\documentclass[reprint,amsmath, amssymb, prd]{revtex4-2}   
\usepackage[utf8]{inputenc}
\usepackage{mwe,float}
\usepackage{caption,subcaption, url}
\usepackage{natbib}
\usepackage[nointegrals]{wasysym}
\usepackage{rotating}
\usepackage{array,tabu,multirow}
\usepackage{graphicx,bm}
\usepackage{dcolumn}
 \usepackage{floatpag} 
 \rotfloatpagestyle{empty}
\usepackage{amsmath}
\usepackage{amsthm}
\usepackage{amssymb}
\usepackage{latexsym,upgreek}
\usepackage{hyperref,rotating}
\usepackage{xpatch,bigints}
\usepackage{booktabs,subcaption}
\usepackage[pdftex]{color}
\usepackage[dvipsnames]{xcolor}
\usepackage{ulem}
\makeatletter
\AtBeginDocument{\xpatchcmd{\@thm}{\thm@headpunct{.}}{\thm@headpunct{\enskip}}{}{}}
\makeatother 
\usepackage{gensymb}
\usepackage{mathdots}

\newcommand{\qand}{\quad \mbox{and} \quad}

\usepackage[version=4]{mhchem}

\usepackage{graphicx,kecpm,slashed}
\usepackage{makeidx}
\makeindex

\newcommand{\bex}{\begin{example}}
\newcommand{\eex}{\end{example}}
\newcommand{\besp}{\begin{split}}
\newcommand{\ensp}{\end{split}}
\newcommand{\ph}{\phantom}
\renewcommand{\thefootnote}{\fnsymbol{footnote}}

\newcommand{\La}{\Lambda}

\newcommand{\om}{\omega}

\newcommand{\by}{\times}

\newcommand{\bos}{\boldsymbol}

\newcommand{\tit}{\textit}

\newcommand{\mcl}{\mathcal}

\newcommand{\btab}{\begin{tabular}}
\newcommand{\etab}{\end{tabular}}
\newcommand{\barr}{\begin{array}}
\newcommand{\earr}{\end{array}}
\newcommand{\bpm}{\begin{pmatrix}}
\newcommand{\epm}{\end{pmatrix}}
\newcommand{\bit}{\begin{itemize}}
\newcommand{\eit}{\end{itemize}}
\newcommand{\ben}{\begin{enumerate}}
\newcommand{\een}{\end{enumerate}}
\newcommand{\bct}{\begin{center}}
\newcommand{\ect}{\end{center}}

\newcommand{\ra}{\rangle}
\newcommand{\la}{\langle}
\newcommand{\bes}{\begin{split}}
\newcommand{\ens}{\end{split}}

\newcommand{\lt}{\left}
\newcommand{\rt}{\right}

\begin{document} 
\title{Is the local Lorentz invariance of general relativity implemented by gauge bosons that have their own Yang-Mills-like action?}

\author{Kevin Cahill}
\email{cahill@unm.edu}

\affiliation{Department of Physics and Astronomy\\
University of New Mexico\\
Albuquerque, New Mexico 87131}
\date{\today}


\begin{abstract}
General relativity with fermions has two independent symmetries: general coordinate invariance and local Lorentz invariance.  General coordinate invariance is implemented by the Levi-Civita connection and by Cartan's tetrads both of which have as their action the Einstein-Hilbert action.  It is suggested here that local Lorentz invariance is implemented not by a combination of the Levi-Civita connection and Cartan's tetrads known as the spin connection, but by independent Lorentz bosons $L^{ab}_{\phantom{ab} i}$ that gauge the Lorentz group, that couple to fermions like Yang-Mills fields, and that have their own Yang-Mills-like action.  A nonsingular $4 \by 4$ hermitian scalar field $h$ is needed to make
the action of the Lorentz bosons invariant under 
local Lorentz transformations.  Lorentz bosons couple to fermion number and generate a spin-dependent static potential that violates the weak equivalence principle.  
If a Higgs mechanism makes them massive, then the static potential also violates the inverse-square law.  Experiments put upper bounds on the strength of such a potential for masses $m_L < 20$ eV.  These upper limits imply that Lorentz bosons, if they exist, are nearly stable and contribute to dark matter.
\end{abstract}

\maketitle

\section{Introduction
\label{Introduction sec}}

General relativity with fermions
has two 
independent symmetries:
general coordinate invariance
and local Lorentz invariance.
General coordinate invariance 
is the well-known, defining symmetry 
of general relativity.
It acts on coordinates
and on the world indexes
of tensors
but leaves Dirac and Lorentz indexes
unchanged.
It is implemented by
the Levi-Civita connection
and by Cartan's tetrads.
\par
Local Lorentz invariance is 
a quite different symmetry.
It acts on Dirac and
Lorentz indexes 
but leaves
coordinates and 
world indexes unchanged.
In standard 
formulations~\citep{Utiyama:1956sy,
Kibble:1961ba,Weinberg:1972tetrads,
Deser:1976ay,Parker:2009uva},
the derivative of a Dirac field
is made covariant 
$\lt(\p_i +
\omega_i \rt) \psi$
by a combination
of the Levi-Civita connection
$\Gamma^j_{\ph{j} k i}$ and 
Cartan's tetrads $c^a_{\ph{a} j}$
known as the 
spin connection 
\begin{equation}
\omega_i
={} \teighth \, \om^{ab}_{\ph{ab} i}
\, \bos [ \c_a, \c_b \bos ] 
\label{omega i in terms of gammas intro}
\end{equation}
in which 
\begin{equation}
\om^{ab}_{\ph{ab} i}
={}
c^a_{\ph{a} j} \, c^{ b k} 
\, \Gamma^j_{\ph{j} k i}
+  c^a_{\ph{a} k} \, \p_i c^{b k}, 
\label{omega ab i}
\end{equation}
$a$ and $b$ are Lorentz indexes,
and $i, j, k$ are world indexes.
\par
Because it
acts on Lorentz and Dirac
indexes but leaves 
world indexes and coordinates 
unchanged, 
local Lorentz invariance
is more like an internal symmetry
than like general coordinate invariance.
In theories with 
local Lorentz invariance
and internal symmetry,
the covariant derivative $D_i$
of a vector of Dirac fields $\psi$
has the spin connection $\om_i$ 
and a matrix $A_i$ of Yang-Mills fields
side by side 
\begin{equation}
D_i \psi
={} 
\lt(\p_i +
\omega_i + A_i \rt) \psi .
\label {naive Dirac covariant derivative intro}
\end{equation}
Just as the Yang-Mills 
connection $A_i$ 
is a linear combination
$A_i ={} -i t^\a A^\a_i$
of the matrices $t^\a$ 
that generate
the internal symmetry group,
so too the spin connection
$\om_i$ is a linear combination
$\om_i ={} \teighth 
\om^{ab}_{\ph{ab} i} \,
\bos [ \c_a, \c_b \bos ]$
of the matrices 
${}- i \tfourth \bos [ \c_a, \c_b \bos ]$
that generate the Lorentz group.
\par
So I ask: Does
the independent symmetry of
local Lorentz invariance have
its own, independent gauge field 
\begin{equation}
L_i ={}  \teighth   
\, 
L^{ a b }_{\ph{a b} i}
\, \bos [ \c_a, \c_b \bos ] 
\end{equation}
with its own field strength
$F_{i k} ={}
\bos [ \p_i + L_i, \p_k + L_k \bos ]$
and Yang-Mills-like action 
\begin{equation}
S_L = {} - \frac{1}{4 f^2}\int
\tr \lt(F^\dag_{ik} \, h
\, F^{ik} \, h^{-1} \rt) 
\sqrt{g} \, d^4x ?
\label {action of the L bosons intro}
\end{equation}
This action is made 
suitably
positive and invariant
under the local Lorentz transformations
\begin{equation}
F' = {} D^{-1}(\La)  F D(\La)
\qand
h ' = {} D^{\dag}(\La) h D(\La)
\end{equation} 
by a
nonsingular $4 \by 4$ hermitian scalar field 
$h$~\citep{Cahill:1978ps,*Cahill:1979qt,*Cahill:1980,*Cahill:1981rq}
whose action density is
\begin{equation}
S_h ={} - M^2
\tr\big[
(D_i h) h^{-1} (D^i h) h^{-1}
\big] 
\label{action of h}
\end{equation} 
in which its covariant
derivative is~\citep{Cahill:1978ps,*Cahill:1979qt,*Cahill:1980,*Cahill:1981rq}
\begin{equation}
D_i h ={}
\p_i h - h L_i  - L_i^\dag h .  
\label{D_i h}
\end{equation} 
The covariant derivative
of a Dirac field would then be
\begin{equation}
D_i \psi ={}
\lt(\p_i  + \teighth 
\, 
L^{ a b }_{\ph{a b} i}
\, \bos{[} \c_a, \c_b \bos ] 
\rt) \psi 
\label {Dirac covariant derivative intro}
\end{equation}
instead of 
the standard form
(\ref{omega i in terms of gammas intro}--\ref
{naive Dirac covariant derivative intro})
\begin{equation}
D_i\psi = \Big(\p_i \,
+ \, \teighth
\big(c^a_{\ph{a} j} \, c^{ b k} 
\, \Gamma^j_{\ph{j} k i}
+  c^a_{\ph{a} k} \, \p_i c^{b k} 
\big)
\bos [ \c_a, \c_b \bos ]
 \Big) \psi.
\end{equation}
If so, then
the Lorentz bosons $L^{a b}_{\ph{ab} i}$ 
couple to fermion number 
and not to mass and  
lead to Coulomb and Yukawa
potentials that violate
the inverse-square law
and the weak equivalence principle.
The hermitian scalar field 
$h$ must assume a nonzero
value in the vacuum
$h_0 ={} \la 0 | h | 0 \ra$
because it is a nonsingular matrix.
It may play the role of a new Higgs field.
Its action (\ref{action of h}) introduces
a new mass into the theory.

\par
Experiments~\citep{Harris:2000zz,
Chen:2014oda,
Lee:2020zjt, Tan:2020vpf,Berge:2017ovy,Tan:2016vwu,
SQYang2012,Lee:2020zjt,Adelberger:2009zz,
Geraci:2008hb,Kapner:2006si,
Smullin:2005iv,Hoyle:2004cw,
LongChan2003,Chiaverini:2002cb,
Lee:2020zjt,Hoskins:1985tn,
Williams:2004qba,
Adelberger:2003zx,
Moody:1993ir,
Hoskins:1985tn,
Spero:1980zz,
Schlamminger:2007ht,
Decca:2005qz,Chen:2014oda,
Chiaverini:2002cb,Geraci:2008hb,
LongChan2003,Tu:2007zz,
Yang:2012zzb,Tan:2016vwu,
Fischbach:1999bc} 
have set upper limits on the 
strength of 
such Yukawa potentials
for Lorentz bosons of mass
less than $20\,$eV\@.
These upper limits imply
that Lorentz bosons,
if they exist, are nearly stable and
contribute to dark matter.
Whether fermions couple 
to Lorentz bosons $L_i$ with 
their own action $S_L$
or to the spin connection $\om_i$
is an open experimental question.
\par 
This paper outlines
a version of general relativity
with fermions in which
the six vector bosons
of the spin connection
$\omega^{ a b }_{\ph{a b} i}$
are replaced by six vector bosons
$L^{ a b }_{\ph{a b} i}$ 
that gauge the Lorentz group and
have their own 
Yang-Mills-like action.
The theory is invariant
under general coordinate
transformations and 
independently under
local Lorentz transformations. 
\par
Section~\ref{General relativity with fermions sec}
sketches the traditional way 
of including fermions 
in a theory of general relativity.
Section~\ref{Local Lorentz Invariance sec}
describes the local Lorentz invariance
of a theory with Lorentz bosons.
Section~\ref{Local Lorentz Invariance 
and Invariance Under
General Coordinate Transformations
Are Independent Symmetries sec}
says why general-coordinate
invariance and local-Lorentz 
invariance are independent
symmetries. 
Section~\ref{Action of the Gauge Fields Omega sec}
describes the Yang-Mills-like action 
(\ref{action of the L bosons intro})
of the gauge fields 
$L^{ a b }_{\ph{a b} i}$
of the local Lorentz group.
Sections~\ref{Making general relativity more similar to gauge theory sec}
and \ref{Making gauge theory more similar to general relativity sec}
suggest ways to make gauge theory and general relativity
more similar to each other.
Section~\ref{Possible Higgs Mechanisms sec}
discusses Higgs mechanisms that
may give masses to the gauge bosons
$L^{ a b }_{\ph{a b} i}$ of the Lorentz group.
Section~\ref{Tests of the Inverse-Square Law sec}
describes some of the constraints
that experimental tests~\citep{Harris:2000zz,
Chen:2014oda,
Lee:2020zjt, Tan:2020vpf,Berge:2017ovy,Tan:2016vwu,
SQYang2012,Lee:2020zjt,Adelberger:2009zz,
Geraci:2008hb,Kapner:2006si,
Smullin:2005iv,Hoyle:2004cw,
LongChan2003,Chiaverini:2002cb,
Lee:2020zjt,Hoskins:1985tn,
Williams:2004qba,
Adelberger:2003zx,
Moody:1993ir,
Hoskins:1985tn,
Spero:1980zz,
Schlamminger:2007ht,
Decca:2005qz,Chen:2014oda,
Chiaverini:2002cb,Geraci:2008hb,
LongChan2003,Tu:2007zz,
Yang:2012zzb,Tan:2016vwu,
Fischbach:1999bc} 
of the inverse-square law
and of the weak equivalence
principle place upon the proposed
theory.
Section~\ref{Lorentz bosons as dark matter sec}
discusses the stability and masses
of $L$ bosons and suggests that they
may be part or all of dark matter.
Section~\ref{Conclusions sec}
summarizes the paper.

\section{General relativity with fermions}
\label{General relativity with fermions sec}

A century ago,
Einstein described gravity
by the action
\begin{equation}
S_E = {}  \frac{1}{16 \pi G}  
\int \! R \, \sqrt{g} \,\, d^4x
= {}  \frac{1}{16 \pi G}  
\int \! g^{ik} \, R_{ik} \, \sqrt{g} \,\, d^4x 
\label {Einstein action}
\end{equation}
in which $ G = {} 1/m_{\text{P}}^2 $
is Newton's constant,
the metric is $(-,+,+,+)$,
letters from the middle
of the alphabet are world indexes,
$g= |\det g_{ik}|$ is the
absolute value of the
determinant of the space-time
metric,
and the Ricci tensor 
$R_{ik} ={} R^\ell_{\ph{\ell} i \ell k}$ 
is the trace of the Riemann tensor
\begin{equation}
R^j_{\phantom{j} i \ell k} ={}
\p_\ell \Gamma^j_{\ph{j} k i} 
- \p_k \Gamma^j_{\ph{j} \ell i}   
+ \Gamma^j_{\ph{j} \ell m} \,
\Gamma^m_{\ph{m} k i} 
- \Gamma^j_{\ph{j} k m} \, 
\Gamma^m_{\ph{m} \ell i}  
\label {Riemann's curvature tensor}
\end{equation}
in which
\begin{equation}
\Gamma^k_{\ph{k} i \ell} ={} 
\thalf \, g^{k j} \left (  \p_\ell g_{j i} 
+ \p_i g_{j \ell} 
- \p_j g_{i \ell} \right)  
= \Gamma^k_{\ph{k} \ell i}
\label{Levi-Civita connection}
\end{equation}
is the Levi-Civita 
connection
which makes the covariant derivative
of the metric vanish~\citep{Bernal:2016lhq}.
\par
The standard action of general relativity 
with fermions is the sum
of the Einstein-Hilbert action 
(\ref{Einstein action})
and the action 
of matter fields
including 
the Dirac action 
\begin{align}
\int - \bar \psi 
\gamma^a c^i_a  
\lt( \p_i + \om_i + A_i
\rt) \psi \, \sqrt{g} \, d^4x .
\label{naive Dirac action om A II}
\end{align}
\par
In what follows, it is proposed
to replace the spin connection
$\om_i$ in the standard Dirac 
action (\ref{naive Dirac action om A II})
with an independent gauge field
$L_i ={} - \teighth \, 
L^{ a b }_{\ph{a b} i}
\, \bos{[} 
\gamma_a, \gamma_b \bos ] $
that has its own action
(\ref{action of the L bosons intro})
and to use 
\begin{align}
S_D ={}&
\int - \bar \psi \,
\gamma^a c^i_a  
\lt( \p_i + L_i + A_i
\rt)   \psi \, \sqrt{g} \, d^4x 
\label {new Dirac action II}
\end{align}
as the action 
of a Dirac field.
This change 
reflects
the independence
of general coordinate invariance
and local Lorentz invariance
and makes
general relativity and
quantum field theory
somewhat more similar.

\section{Local Lorentz Invariance}
\label{Local Lorentz Invariance sec}

The Einstein action (\ref{Einstein action})
has a trivial 
symmetry under 
local Lorentz transformations
that act on Lorentz indexes
but
leave world indexes 
and coordinates 
unchanged.
This symmetry
becomes apparent 
when Cartan's tetrads $c^a_{\ph{a} i}$ and
$c_{\ph{b} k}^b$ are used
to write the metric $g_{ik}$ 
in a form
\begin{equation}
g_{ik}(x) ={} c^a_{\ph{a} i}(x) 
 \, \eta_{ab} \, c_{\ph{b} k}^b(x)
\label{metric}
\end{equation}
that is unchanged by 
local Lorentz transformations
\begin{equation}
c'^a_{\ph{' a} i}(x) ={} 
\La^a_{\ph{a} b}(x) \, c_{\ph{b} i}^b(x).
\label{local Lorentz transformation for tetrads}
\end{equation}
The Levi-Civita connection
(\ref{Levi-Civita connection}) and
the action (\ref{Einstein action})
are defined in terms of the metric
and so 
are also invariant under
local Lorentz transformations.
\par
More importantly,
the two Dirac actions 
(\ref{naive Dirac action om A II})
and (\ref{new Dirac action II}) 
have a nontrivial 
symmetry under 
local Lorentz transformations.
Under such a 
local Lorentz transformation,
a Dirac field
transforms under the
$(\thalf,0) \oplus (0, \thalf)$
representation $D(\La)$
of the Lorentz group
with no change in 
its coordinates $x$
\begin{equation}
\psi'_\a(x) ={} D^{-1}_{\a \b}(\La(x)) 
\, \psi_\b(x) .
\label{local Lorentz transformation of psi}
\end{equation}
The Lorentz-boson matrix 
$ L_i ={} - \teighth 
\, 
L^{ a b }_{\ph{a b} i}
\, \bos [ \c_a, \c_b \bos ] $ makes
$\p_i + L_i$ a covariant
derivative   
\begin{equation}
\p_i + L'_i = {}
D^{-1} (\Lambda) ( \p_i + L_i ) 
D(\Lambda).
\label{how covariant derivative goes}
\end{equation}
In more detail 
with $\La =\La(x)$, 
the matrix $L_i$ transforms as
\begin{align}
L'_i = {}&
D^{-1}(\Lambda) 
\, \p_i D(\Lambda) 
+
D^{-1}(\Lambda) \, L_i \, D(\Lambda)
\nn\\
={}&
D^{-1}(\Lambda) 
\, \p_i D(\Lambda) 
- \teighth \,
D^{-1}(\Lambda) \,
L^{ a b }_{\ph{a b} i}
\, \bos{[} \c_a, \c_b \bos ] 
\, D(\Lambda)
\label{how Li goes}
\nn\\
={}&
D^{-1}(\Lambda) 
\, \p_i D(\Lambda) 
- \teighth \, 
L^{ a b }_{\ph{a b} i}
\, \La_a^{\ph{a} c} \La_b^{\ph{b} d}
\, \bos{[} \c_c, \c_d \bos ] 
\end{align}
in which 
$\La_a^{\ph{a} c} =
\La^{-1 c}_{\ph{-1 c} a}$.
Since 
\begin{equation}
\tr \lt( \bos[ \c^a, \c^b \bos] 
\bos[ \c_c, \c_d \bos] \rt)
={}
16 \lt( \d^a_d \d^b_c 
- \d^a_c \d^b_d 
\rt) ,
\end{equation}
its components transform as
\begin{align}
L'^{ab}_{\ph{ab} i} 
={}&
- \thalf \, \tr ( L'_i \bos [ \c^a, \c^b \bos ])
\\
={}&
\La_c^{\ph{c} a} \La_d^{\ph{d} b} 
L^{cd}_{\ph{cd} i} 
- \thalf \, \tr \lt( D^{-1}(\Lambda) 
\p_i D(\Lambda) \bos [ \c^a, \c^b \bos ] \rt).\nn
\end{align}
Under an infinitesimal transformation
\begin{equation}
\La = I + \l
\qand
D(\l) = I - \teighth \l_{ab} 
\bos [ \c^a, \c^b \bos ],
\label{infinitesimal transformation}
\end{equation}
the Lorentz bosons transform as
\begin{equation}
L'^{ab}_{\ph{ab} i} 
={}
L^{ab}_{\ph{ab} i} 
+
L^{cb}_{\ph{cb} i} \l_c^{\ph{c} a}
+
L^{ad}_{\ph{ad} i} \l_d^{\ph{d} b}
+
\p_i \l^{ab}.
\label{Lorentz bosons transform infinitesimally}
\end{equation}
The components of the spin connection
obey similar equations, and 
the conventional Dirac action 
(\ref{naive Dirac action om A II})
also is
invariant under local Lorentz transformations. 
\par
Local Lorentz transformations
operate
on the Lorentz indexes $a, b, c, \dots$
of the tetrads, of the
spin connection $\om^{a b}_{\ph{a b} i}$,
and of the gamma matrices 
$\c^a \bos [ \c_b, \c_c \bos ]$,
and also on the Dirac indexes $\a, \b, \c$
of the gamma matrices and
of the Dirac fields $\bar \psi, \psi$,
but not upon the world index $i$ or
the spacetime coordinates $x$.
In this sense,
the invariance of Dirac's action $S_D$
under local Lorentz transformations
is like an internal symmetry.

\section{Local Lorentz Invariance 
and Invariance Under
General Coordinate Transformations
Are Independent Symmetries}
\label{Local Lorentz Invariance 
and Invariance Under
General Coordinate Transformations
Are Independent Symmetries sec}

\par
The Dirac action $S_D$
is invariant both under
a local Lorentz 
transformation $\La(x)$ and under a  
general coordinate transformation
$ x \to x'$.
Under a local Lorentz 
transformation $\La(x)$,
the coordinates are unchanged,
$x'=x$, and
the fields transform as
\begin{align}
\psi'_\a(x) ={} &
D^{-1}_{\a \b}(\La(x)) \, \psi_\b(x)
\nn\\
c'^a_{\ph{' a} i}(x)  ={}&  
\La^a_{\ph{i} b}(x) \,
c^b_{\phantom{b} i}(x)
\nn\\
L'^{ab}_{\ph{'ab} i}(x) ={}&
\La_c^{\ph{c} a}(x) \La_d^{\ph{d} b}(x)
L^{cd}_{\ph{cd} i}(x) 
\nn\\
- {}&
\thalf  \tr \lt( D^{-1}(\Lambda(x)) 
\p_i D(\Lambda(x))
\bos [ \c^a, \c^b \bos ] \rt)
\nn\\
h'(x) ={}&  D^\dag(\Lambda(x))
h(x) D(\Lambda(x))
\label{tetrads and spin connection transform}
\end{align}
in which $\La_c^{\ph{c} a}(x) ={}
\La^{-1 a}_{\ph{-1 a} c}$.
Under a general coordinate
transformation, the fields transform as
\begin{align}
\psi'_\a(x') ={} & \psi_\a(x)
\nn\\
c'^a_{\phantom{' a} i}(x')  ={}&  
\frac{\p x^k}{\p x'^i} \, 
c^a_{\phantom{b} k}(x)
\nn\\
L'^{ab}_{\ph{ab} i}(x') ={}&
\frac{\p x^k}{\p x'^i} \,
L^{ab}_{\ph{ab} k}(x) 
\nn\\
h(x') ={}&  h(x) .
\label{tetrads and spin connection transform 2}
\end{align}
The two transformations,
$\psi_\a(x) 
\to D^{-1}_{\a \b}(\La(x)) 
\, \psi_\b(x)$
and $x \to x'$,
are different and independent;
the
coordinates $x'$ and
$\La(x) x$ are unrelated.
\par
Every conventional, local Lorentz
transformation is a 
general coordinate transformation, 
so one might be tempted 
to imagine that every 
general coordinate transformation
is a conventional, local  Lorentz
transformation.
But one can see that this is not 
the case by comparing the
infinitesimal form of a
general coordinate transformation
\begin{equation}
dx'^i ={} \frac{\p x'^i }{\p x^k}
\, dx^k
\label{general coordinate transformation}
\end{equation}
which has 16 generators
with that of a conventional, local Lorentz
transformation
\begin{equation}
dx'^a ={} \La^a_{\ph{a} b}
\, dx^b  
=
dx^a + (\bos \ep_r \cdot \bos R^a_{\ph{a} b}
+ \bos \ep_b \cdot \bos B^a_{\ph{a} b}) \, dx^b
\label{tiny Lorentz transformation}
\end{equation}
which has only 
6~\citep{CahillCUP2lorentzgroup}.
\par
Special relativity offers
another temptation.
In special relativity,
global Lorentz transformations
$\La$ act on the
spacetime coordinates
and on the indexes
of a Dirac field
\begin{equation}
\begin{split}
x'^a ={}& \La^a_{\ph{a} b} x^b
\\
\psi'_\a(x') ={}& D^{-1}_{\a \b}(\La) 
\, \psi_\b(\La x) .
\label{SR Lorentz transformation of psi}
\end{split}
\end{equation}
This global Lorentz transformation
leaves
the specially relativistic
Dirac action density 
unchanged
\begin{align}
\big[- i \psi^\dag
\c^0 \gamma^a  
\p_a  \psi \big]' 
={}&
- i \psi^\dag D^{-1 \dag} 
\c^0 \c^a  
D^{-1} \p'_a  \psi
\nn\\
={}&
- i \psi^\dag 
\c^0 D \c^a  
D^{-1} \La_a^{\ph{a} c}\p_c  \psi
\nn\\
={}&
- i \psi^\dag 
\c^0 
\c^b \Lambda_b^{\ph{b} a}
\La_a^{\ph{a} c}\p_c  \psi
\label{SR Dirac action unchanged}
\\
={}&
- i \psi^\dag 
\c^0 
\c^b \d^c_b \p_c  \psi
= - i \psi^\dag 
\c^0 
\c^b  \p_b  \psi. 
\nn
\end{align} 
\par
But in general relativity 
with fermions, 
Cartan's tetrads $c_a^i$
allow the action to be invariant
under a local Lorentz transformation
without a corresponding general
coordinate transformation.
The matrix $D^{-1}_{\a \b}(\La(x))$
represents a local Lorentz transformation
and acts
(\ref{local Lorentz transformation of psi})
on the spinor indexes
of the Dirac field but not on
its spacetime coordinates.
Since 
\begin{equation}
D \c^a D^{-1} = {}
\La_{a'}^{\ph{a'} a} \c^{a'}
\qand
D \c_a D^{-1} = {}
\La^{a'}_{\ph{a'} a} \c_{a'},
\end{equation}
a local Lorentz transformation
(\ref{how Li goes})
does not change 
\begin{equation}
D \gamma^a D^{-1} c'^i_a =
\c^b \Lambda_b^{\ph{b} a}
\La_a^{\ph{a} c} c_c^i
= \c^b \d^c_b c_c^i
= \c^b c_b^i.
\end{equation}
But the effect of a
local Lorentz transformation
(\ref{how Li goes})
on the Lorentz matrix $L_i$ is
\begin{align}
D (\p_i + L'_i )D^{-1} &  - \p_i 
={}
- \teighth 
L'^{ a b }_{\ph{'a b} i}
\, D \bos{[}      
\c_a, \c_b \bos ] D^{-1}
\nn\\
={}& - \teighth 
\Lambda^{\ph{c} a}_c 
\Lambda^{\ph{d} b}_d 
L^{c d}_{\ph{a b} i}
\, D  \bos{[}        
\c_a, \c_b \bos ] D^{-1} 
\nn\\
={}&
- \teighth 
\Lambda^{-1 a}_{\ph{-1 a} c}
\Lambda^{-1 b}_{\ph{-1 b} d}
L^{c d}_{\ph{a b} i}
\La^e_{\ph{a} a}
\La^f_{\ph{a} b}
 \bos [ \c_e, \c_f \bos ] 
\nn\\
={}&
- \teighth 
\d^e_{\ph{e} c}
\d^f_{\ph{f} d} 
L^{c d}_{\ph{a b} i}
\bos [ \c_e, \c_f \bos ] 
\nn\\
={}&
- \teighth 
L^{ c d }_{\ph{a b} i}
\, \bos{[}      
\c_c, \c_d \bos ] 
= L_i 
\label{local Lorentz transformation of L}
\end{align}
so that
\begin{equation}
D (\p_i + L'_i )D^{-1}
={}
\p_i + L_i .
\label{Lorentz transformation on Lorentz matrix}
\end{equation}
A local Lorentz transformation
therefore leaves the
Dirac action density
invariant
\begin{align}
\big[- i \psi^\dag 
& \c^0 
\gamma^a c^i_a  
\lt( \p_i + L_i 
\rt)  \psi \big]'
\nn\\
={}& 
- i \psi^\dag D^{-1 \dag}\c^0 
\gamma^a c'^i_a  
\lt( \p_i + L'_i 
\rt)  D^{-1} \psi
\nn\\
={}& 
- i \psi^\dag D^{-1 \dag}\c^0 
\gamma^a c'^i_a  D^{-1} D
\lt( \p_i + L'_i 
\rt)  D^{-1} \psi
\nn\\
={}& 
- i \psi^\dag \c^0 
D
\gamma^a c'^i_a  
D^{-1} \! \lt( \p_i + L_i \rt) 
\psi
\\
={}&
- i \psi^\dag \c^0 
\gamma^a c^i_a 
\lt( \p_i +  L_i 
\rt)   \psi.
\nn
\end{align}
\par
The symmetry under local 
Lorentz transformations
is independent of 
the symmetry under
general coordinate
transformations.
They are independent
symmetries.  

\section{Action of the Gauge Fields $L_i$}
\label{Action of the Gauge Fields Omega sec}
\par
Since local Lorentz symmetry
is like an internal symmetry,
its gauge fields 
$L_i ={}  \teighth 
L^{ a b }_{\ph{a b} i}
\, \bos{[} 
\gamma_a, \gamma_b \bos ] $ 
should have an action 
like that of a Yang-Mills field
\begin{equation}
S_L = {} - \frac{1}{4 f^2}\int
\tr \lt(F^\dag_{ik} \, {h} 
\, F^{ik} \, {h^{-1}} \rt) 
\sqrt{g} \, d^4x 
\label{action of the Lorentz connection}
\end{equation}
in which
\begin{equation}
F_{i k} ={}
\bos [ \p_i + L_i, \p_k + L_k \bos ] ,
\label{F is}
\end{equation}
and $h$ is a  $4 \by 4 $
nonsingular hermitian 
matrix~\citep{Cahill:1978ps,*Cahill:1979qt,*Cahill:1980,*Cahill:1981rq} .
Under local Lorentz transformations
$ D(\Lambda(x)) $,
the field strengths $F^{ik}$
and the matrix $h$
transform as
\begin{equation}
\begin{split}
F'^{ik}(x) ={} &
D^{-1}(\Lambda(x))
F^{ik}(x) D(\Lambda(x))
\\
h'(x) ={}& D^\dag(\Lambda(x))
h(x) D(\Lambda(x))
\label{how F and h go}
\end{split}
\end{equation}
and so
the action $S_L$ is invariant
under local Lorentz transformations
as well as under 
general coordinate 
transformations~\citep{Cahill:1978ps,*Cahill:1979qt,*Cahill:1980,*Cahill:1981rq} .
\par
One might think 
that it would be sufficient 
to set $h = \b = i \c^0$,
since $D^\dag (\Lambda) \b D(\Lambda)
= {} \b$ 
and
$D(\Lambda) \b D^\dag (\Lambda) = \b$,
but the resulting action
$S_L$ is not bounded below.
\par
In terms of the gamma matrices
\begin{equation}
\begin{split}
\c^0 = {}& -i \, \bpm
0 & 1 \\
1 & 0 \\
\epm,
\quad
\c^i = - i \, 
\bpm
0 & \s^i \\
{} - \s^i & 0 \\
\epm ,
\\{}&
\qand
\c^5 =
\begin{pmatrix}
1 & 0 \\
0 & -1 \\
\end{pmatrix},
\label {Weinberg's Dirac matrices} 
\end{split}
\end{equation}
the commutators in 
$L_i ={} - \teighth 
L^{ a b }_{\ph{a b} i}
\, \bos{[} 
\gamma_a, \gamma_b \bos ] $
are for spatial $a,b,c=1,2,3$,
\begin{equation}
\boldsymbol{[} 
\gamma_a, \gamma_b \bos ]
= 2i \ep_{abc} \s^c I
\qand
\bos[ \c_0, \c_a \bos]
= -2 \s^a \c^5.
\end{equation}
So setting
\begin{equation}
    \bos r^a_{\ph{s} i} ={}\thalf \ep_{a b c}
    L^{bc}_{\ph{bc} i}
\qand
\bos b^a_{\ph{a} i} = {} 
L^{a0}_{\ph{a0} i} ,
\label{rotons and boostals}
\end{equation}
the matrix of gauge fields 
$L_i$ is
\begin{align}
L_i ={}& 
- \teighth \, 
L^{ a b }_{\ph{a b} i}
\, \bos [ 
\gamma_a, \gamma_b \bos ]
= 
-i \, \thalf \, \bos r_i \cdot \bos \s \, I
- \thalf \, \bos b_i \cdot \bos \s 
\, \c^5 ,
\label{L is}
\end{align}
and its field strength 
(\ref{F is})
is
\begin{align}
F_{i k} ={}& 
\bos [ \p_i + L_i, \p_k + L_k \bos ] 
\\
={}&
-i \, \thalf \, 
(\p_i \bos r_k - \p_k \bos r_i
+ \bos r_i \times \bos r_k 
- \bos b_i \by \bos b_k
) 
\cdot \bos \s \, I
\nn\\
{}& - \, \thalf \, ( \p_i \bos b_k - \p_k \bos b_i 
+ \bos r_i \by \bos b_k 
+ \bos b_i \by \bos r_k) 
\cdot \bos \s \, \c^5 .
\nn
\end{align}
With the abbreviations
\begin{equation}
\begin{split}
\bos R_{ik} ={}&
(\p_i \bos r_k - \p_k \bos r_i
+ \bos r_i \times \bos r_k 
- \bos b_i \by \bos b_k
) 
\\
\bos B_{ik} =&
( \p_i \bos b_k - \p_k \bos b_i 
+ \bos r_i \by \bos b_k 
+ \bos b_i \by \bos r_k) ,
\end{split}
\end{equation}
the action $S_L$ is 
the trace
\begin{equation}
\begin{split}
S_L ={}&
- \frac{1}{16f^2} \int \tr
\Big[\big( \bos R_{ik} \cdot \bos \s I 
+i  \bos B_{ik} \cdot \bos \s \c^5
\big) h
\\
{}&
\qquad 
\big( \bos R^{ik} \cdot \bos \s I 
- i \bos B^{ik} \cdot \bos \s \c^5
\big) h^{-1}
\Big]  \sqrt{g} \, d^4x .
\end{split}
\end{equation}
\par
The $4 \by 4 $ nonsingular
hermitian matrix $h$ 
may play a role in the
action (\ref{new Dirac action II})
of a spin-one-half field $\psi$
because
we may take the quantity
$\bar \psi$ either to be
$\bar \psi = \psi^\dag \b$
as usual or to be 
$\bar \psi = \psi^\dag h$ .
The covariant derivative
(\ref{D_i h}) of $h$  
transforms as
$ \big( D_i h \big)' ={}
D^{\dag}(\La) \, D_i h \, D(\La) $ .

\section{Making general relativity more similar to gauge theory}
\label{Making general relativity more similar to gauge theory sec}

There are three reasons to define 
the covariant derivative
of a Dirac field 
in terms of Lorentz bosons
\begin{equation}
L_i ={} 
 \teighth \, 
L^{ a b }_{\ph{a b} i}
\, \bos [ \c_a, \c_b \bos ]
\label{Lorentz connection}
\end{equation}
with their own action
(\ref{action of the Lorentz connection}) 
as
\begin{equation}
D_i \psi = {} \lt( \p_i + L_i + A_i
\rt)   \psi 
\label{Dirac covariant derivative}
\end{equation}
rather than in terms 
of the spin connection
(\ref{omega i in terms of gammas intro})
\begin{equation}
\begin{split}
\om_i ={}&
\teighth         
\om^{ a b }_{\ph{a b} i}
\, \bos [ 
\c_a, \c_b \bos ]
\\
= {}&
\teighth         
\lt(c^a_{\ph{a} j} \, c^{b k} 
\, \Gamma^j_{\ph{j} k i}
+  c^a_{\ph{a} k} \, \p_i c^{b k}
\rt)
\, \bos [ \c_a, \c_b \bos ]
\label{omega in terms of Gamma and tetrads}
\end{split}
\end{equation}
as (\ref{naive Dirac covariant derivative intro})
\begin{equation}
\lt( \p_i + \om_i + A_i
\rt)   \psi .
\label{spin-connection covariant derivative}
\end{equation}

\par
One reason is that 
the symmetry of local Lorentz
transformations is independent
of the symmetry of
general coordinate
transformations. 
So local Lorentz invariance
should have its own gauge field
$L_i$ and action $S_L$
independent of the tetrads 
and the Levi-Civita connection
of general coordinate
transformations.
\par
A second reason to prefer
the Lorentz connection $L_i$
to the spin connection $\om_i$
is that the $L$-boson 
covariant derivative
\begin{equation}
\lt(\p_i  + \teighth 
\, 
L^{ a b }_{\ph{a b} i}
\, \bos{[} \c_a, \c_b \bos ] 
\rt) \psi 
\end{equation}
is simpler than 
the spin-connection
covariant derivative
\begin{align}
\Big( \p_i + \teighth {}         
( 
c^a_{\ph{a} j} \, c^{b k} 
\, \Gamma^j_{\ph{j} k i}
+  c^a_{\ph{a} k} \, \p_i c^{b k}
 ) \,
\boldsymbol{[} 
\gamma_a, \gamma_b \bos ]
\Big) \, \psi .
\label{ugly Dirac action}
\end{align}
\par
A third reason
is that using the Lorentz connection
(\ref{Lorentz connection}),
the Dirac covariant derivative
(\ref{Dirac covariant derivative}), and
the action 
(\ref{action of the Lorentz connection}) 
for the Lorentz connection,
makes general relativity
with fermions more similar 
to the gauge theories 
 of the
standard model.

\section{Making gauge theory more similar to general relativity}
\label{Making gauge theory more similar to general relativity sec}

Under a local Lorentz transformation,
the spin connection $\om_i$ changes
more naturally, more
automatically than does the Lorentz
connection $L_i$.
The automatic feature of
the spin connection is that
its definition (\ref{omega ab i})
implies that under infinitesimal
(\ref{infinitesimal transformation})
and finite local Lorentz transformations
it transforms as
\begin{equation}
\om'^{ab}_{\ph{'ab} i}
={}
\om^{ab}_{\ph{ab} i} 
+
\om^{cb}_{\ph{cb} i} \l_c^{\ph{c} a}
+
\om^{ad}_{\ph{ad} i} \l_d^{\ph{d} b}
+
\p_i \l^{ab}
\label{how omega goes infinitesimally}
\end{equation}
and as
\begin{equation}
\begin{split}
\om'^{ab}_{\ph{'ab} i}
={}&
\La_c^{\ph{c} a} \La_d^{\ph{d} b} 
\om^{cd}_{\ph{ab} i} +
\La^{\ph{c} a}_c \p_i \La^{cb}.
\label{how omega goes grandly}
\end{split}
\end{equation}
The terms $\p_i \l^{ab}$ and
$\La^{\ph{c} a}_c \p_i \La^{cb}$
occur automatically
without the need to put in 
by hand a term like
$D^{-1} (\Lambda) \p_i D(\Lambda)$.
\par
Terms like
$D^{-1} (\Lambda) \, \p_i D(\Lambda)$
are a common feature of 
gauge theories whether abelian
or nonabelian.  
We can make them occur 
automatically
in local Lorentz transformations
if we add to the
Lorentz connection 
$L^{ a b }_{\ph{'a b} i}$ 
the term 
$u^a_{\ph{a} \a} \, \p_i \, u^{b \a}$
in which the four
Lorentz
4-vectors $u^{a \a}(x)$
obey the condition
\begin{equation}
u^{a\a} \eta_{\a \b} u^{b\b} = 
{} \eta^{ab},
\label{Minkowski condition}
\end{equation}
and $\a = 0, 1, 2, 3$ is a label,
not an index.
It follows then from
this condition 
(\ref{Minkowski condition})
on the quartet of vectors $u^{a\a}$
that the augmented Lorentz connection
$L^{a b}_{\text{new} \, i}$
\begin{equation}
L^{ a b }_{\text{new} \, i}
={}
L^{ab}_{\ph{cd} i} 
+ u^a_{\ph{a} \a} \, \p_i \, u^{b \a}
\end{equation}
automatically changes under 
a local Lorentz transformation
$\La_c^{\ph{c} a}
={} \La^{-1 a}_{\ph{-1 a} c}$ to
\begin{align}
L'^{ a b }_{\text{new} \, i}
={}&
\La_c^{\ph{c} a} \La_d^{\ph{d} b}
\, L^{c d}_{\ph{cd} \, i} 
+ \La_c^{\ph{c} a}
u^c_{\ph{a} \a} \, \p_i \, 
(\La_d^{\ph{d} b} u^{d \a})
\nn\\
={}& 
\La_c^{\ph{c} a} \La_d^{\ph{d} b}
\, ( L^{c d}_{\ph{cd} \, i}
+ u^c_{\ph{a} \a} \, \p_i \, u^{d \a} )
+ u^c_{\ph{a} \a} u^{d \a}
\La_c^{\ph{c} a}
 \, \p_i \, 
\La_d^{\ph{d} b}
\nn\\
={}& 
\La_c^{\ph{c} a} \La_d^{\ph{d} b}
\, L^{c d}_{\text{new} \, i}
+ \eta^{cd}
\La_c^{\ph{c} a}
 \, \p_i \, 
\La_d^{\ph{d} b} 
\nn\\
={}& 
\La_c^{\ph{c} a} \La_d^{\ph{d} b}
\, L^{c d}_{\text{new} \, i}
+ 
\La_c^{\ph{c} a}
 \, \p_i \, 
\La^{c b} 
\nn\\
={}& 
\La_c^{\ph{c} a} \La_d^{\ph{d} b}
\, L^{c d}_{\text{new} \, i}
+ 
\La^{-1 a}_{\ph{-1 a} c}
 \, \p_i \, 
\La^{c b}
\label{L automatically changes}
\end{align}
without the need to explicitly
add the last term
$\La^{-1 a}_{\ph{-1 a} c}
 \, \p_i \, 
\La^{c b}$
by hand.
\par
In matrix form,
the condition (\ref{Minkowski condition}) 
is the requirement 
\begin{equation}
u^{a\a} \eta_{\a \b} u^{b\b} = 
{} \eta^{ab}
\end{equation}
that the matrix formed 
by the quartet
of vectors $u^{a \a}$ 
be a Lorentz transformation
\begin{equation}
x_a u^{a \a}  \eta_{\a \b}
u^{b \b} y_b
={} x_a \eta^{ab} y_b .
\end{equation}
\par
The augmentation of the 
Lorentz connection
$L^{a b}_{\ph{ab} i} \to 
L^{a b}_{\text{new} \, i}$
by the addition of the term
$u^a_{\ph{a} \a} \, \p_i \, u^{b \a}$,
which is similar to the tetrad term
$c^a_{\ph{a} k} \, \p_i c^{b k}$
of the spin connection
(\ref{omega ab i}),
makes its change 
(\ref{L automatically changes}) 
under
local Lorentz transformations
as automatic as that
(\ref{how omega goes grandly})
of the spin connection.
\par
The use of a more automatic 
connection makes
gauge theory more similar 
to general relativity
with fermions.
\par
We can extend the use of such terms
to internal symmetries and so
make the inhomogeneous terms
appear automatically rather than
by hand or by fiat.
For instance, we can augment
the abelian connection $A_i$ to
\begin{equation}
A_{\text{new} \, i}(x) = {}
A_i(x) + e^{-i \phi(x)} \p_i e^{i \phi(x)}
\end{equation}
in which $\phi(x)$ is an arbitrary phase.
A $U(1)$ transformation
\begin{equation}
e^{-i \phi(x)} \to e^{-i (\th(x) + \phi(x))}
\,\,\, \text{and} \,\,\,
\psi(x) \to e^{-i\th(x)} \psi(x)
\end{equation}
would then change
the covariant derivative
$(\p_i + A_{\text{new} \, i}) \psi$
to 
\begin{align}
[(\p_i +& A_{\text{new} \, i}) \psi]'
={} 
(\p_i + A_i 
+  e^{-i (\th + \phi)} 
\p_i e^{i (\th + \phi)}  )  
e^{-i\th}\psi
\nn\\
={}& 
e^{-i\th}(\p_i - i \p_i \th + A_i 
+ i \p_i \th          
+ e^{-i \phi} 
\p_i e^{i \phi}  ) \psi
\\
={}& 
e^{-i\th}(\p_i  + A_i 
+ e^{-i \phi} 
\p_i e^{i \phi}  ) \psi       
= e^{-i\th}         
(\p_i  + A_{\text{new} \, i}) 
\psi .
\nn
\end{align}
\par
Similarly, we can augment 
the nonabelian connection
$A_i = {} -i t^\a A^\a_i$ 
for $SU(n)$ to
\begin{equation}
A_{\text{new} \, i}(x) ={}
A_i(x) + u_{\a \b}(x) 
\p_i u^*_{\a \c}(x)
\end{equation}
in which the $n$ $n$-vectors $u_{\a \c}$
are orthonormal
\begin{equation}
u_{\b \a} u^*_{\c \a} 
= {} \d_{\b \c}.
\end{equation}
An $SU(n)$ transformation
\begin{equation}
A_i \to g A_i g^\dag,
\,\,\,
u \to g u,
\,\,\, \text{and} \,\,\,
\psi \to g \psi
\end{equation}
would then change
the covariant derivative
$(\p_i + A_{\text{new} \, i}) \psi$
to 
\begin{align}
[(\p_i + &A_{\text{new} \, i}) \psi]'
={}
( \p_i + gA_ig^\dag + 
g \, u \, \p_i ( u^\dag g^\dag) ) g \psi
\nn\\
={}&
g ( \p_i + g^\dag \p_i g + A_i  + 
(\p_i g^\dag) u u^\dag g
+ u \, \p_i \, u^\dag  )  \psi
\nn\\
={}&
g ( \p_i + g^\dag \p_i g + A_i  + 
(\p_i g^\dag) g
+ u \, \p_i \, u^\dag  )  \psi
\nn\\
={}&
g ( \p_i + A_i  + 
u \, \p_i \, u^\dag  )  \psi
= g ( \p_i + A_{\text{new} \, i} ) \psi.
\end{align}

\section{Possible Higgs Mechanisms}
\label{Possible Higgs Mechanisms sec}

\par
The $4 \by 4$ nonsingular 
hermitian matrix $h$
is needed to make the action 
(\ref{action of the Lorentz connection})
gauge invariant.  It also may replace $\b$
in the fermionic action
(\ref{new Dirac action II}).
Since it is nonsingular, it
must assume a nonzero average
value in the vacuum
\begin{equation}
h_0 ={} \la 0 | h | 0 \ra .
\end{equation}
So we have a new kind of 
Higgs mechanism that can give 
masses to gauge fields and fermions.
This will be taken up in later papers.
\par
Other kinds of Higgs mechanisms
are also possible with different Higgs fields.
The actions $S_{L}$ and $S_D$ 
(\ref{action of the L bosons intro} \& 
\ref{new Dirac action II})
leave the gauge
bosons $L$ massless,
but a Higgs mechanism is possible.
An interaction with
a field $v^a$ that is a scalar under 
general coordinate transformations but
a vector under local Lorentz transformations
has as its covariant derivative
\begin{equation}
D_i v^a = {} \p_i  v^a + L^a_{\ph{a} b i}
v^b .
\label{covariant derivative of v^a}
\end{equation}
If the time component
$v^0$ has a nonzero mean value
in the vacuum
$\la 0 | v^0| 0 \ra \ne 0$,
then the scalar 
$ - {} D_i v^a D^i v_a $
contains a mass term
\begin{equation}
{} - L^a_{\ph{a} 0 i}
v^0 \, L_a^{\ph{a} 0 i} v_0 
={}
- L^{a 0}_{\ph{a 0} i}
v^0 \, L^{a 0 i} v^0
\label{temporal mass term}
\end{equation}
that makes the boost vector bosons 
$b^s_{\ph{s} i} ={} L^{s0}_{\ph{s0} i}$
massive but leaves the
rotational vector bosons 
$r^s_{\ph{s} i} ={} 
\thalf \ep_{ s t u}L^{tu}_{\ph{tu} i}$
massless.
On the other hand, if 
the spatial components have a
nonzero mean value,
$\la 0 | \bos v|0\ra \ne 0$,
then the mass term is
\begin{equation}
{} - L^a_{\ph{a} s i}
v^s \, L_a^{\ph{a} s i} v_s 
={}
- L^{s' s}_{\ph{s' s}  i}
v^s \, L^{s' s i} v^s 
+ L^{0 s}_{\ph{a s}  i}
v^s \, 
L^{0 s i} v^s .
\label{spatial mass term}
\end{equation}
Adding the two mass terms
(\ref{temporal mass term} and
\ref{spatial mass term}),
we find
\begin{equation}
\begin{split}
{} - L^a_{\ph{a} 0 i}
v^b \, L_a^{\ph{a} 0 i} v_b 
={}&
L^{0 s}_{\ph{0 s}  i}
v^a \,    
L^{0 s i} v_a 
- L^{s' s}_{\ph{s' s}  i}
v^s \, L^{s' s i} v^s 
\label{Adding the two mass terms}
\end{split}
\end{equation}
which makes all six
gauge bosons
massive as long as the
mean value is timelike
\begin{equation}
\la 0 | v^a v_a | 0 \ra < 0
\label{mean value is timelike}
\end{equation}
and at least two spatial
components $\la 0 |v^s|0\ra \ne 0$
have nonzero mean values
in the vacuum.
This condition holds
in all Lorentz frames if
three  vectors 
$v^{a 1}, v^{a 2},$ and $ v^{a 3}$
have different timelike mean values
in the vacuum.
\par
In the vacuum of flat space, 
tetrads have mean values 
that are Lorentz transforms of
$ c^a_i ={} \d^a_i $
and that produce
the Minkowski metric (\ref{metric}) 
\begin{equation}
g_{ik} ={} \d^a_i \, \eta_{a b} \, \d^b_k
= \eta_{ik} .
\label{frozen  metric}
\end{equation}
So it is tempting 
to look for a Higgs mechanism 
that uses the covariant
derivatives $D_\ell \, c^a_{\ph{i} k}$
of the tetrads.
For $\Gamma^j_{\ph{j} k \ell} = 0$
and $c^a_{\ph{a} k} = \d^a_k$,
the term 
\begin{equation}
\begin{split}
- \thalf \, m_L^2 \, 
(D^i \, c_a^{\ph{a} k} )
\, D_i \, c^a_{\ph{a} k} 
={}&  
- \thalf \, m_L^2 \, 
L_a^{\ph{a} b i} \, 
c_b^{\ph{b} k} \,\,
L^a_{\ph{a} c i} \,
c^c_{\ph{b} k}
\\
={}&
- \thalf \, m_L^2 \, 
L_{a b}^{\ph{a b} i} \, 
L^{a b}_{\ph{a b} i} 
\end{split}
\end{equation} 
makes the rotational bosons
$ \bos r^s_{\ph{s} i} = 
\thalf \ep_{ s t u}L^{tu}_{\ph{tu} i}$
massive but makes the boost bosons
$ \bos b^s_{\ph{s} i} = 
{} L^{s0}_{\ph{s0} i} $
tachyons.
If weakly coupled tachyons
are unacceptable, then
the Higgs mechanism 
(\ref{covariant derivative of v^a}--\ref
{Adding the two mass terms})
that uses three 
world-scalar Lorentz vectors
with different time-like mean values
in the vacuum
$ v^a_1 $, $ v^a_2 $, 
$ v^a_3 $
is a more
plausible way to make the gauge
bosons $L^{ab}_{\ph{ab} i}$
massive.

\section{Tests of the Inverse-Square Law}
\label{Tests of the Inverse-Square Law sec}

\par
In the static limit,
the exchange of six
Lorentz bosons $L^{ab}_{\ph{ab} i}$
of mass $m_L$ would 
imply that two macroscopic
bodies of $F$ and $F'$ fermions
separated by a distance $r$
would contribute to the energy
a static Yukawa potential
\begin{equation}
V_L(r) = \frac{3 \, F F' f^2}{2 \pi r}
\, e^{- m_L r} .
\label{would have a static potential}
\end{equation}
This potential is positive and repulsive
(between fermions and between antifermions)
because 
the $L$'s  are vector bosons.
It violates the weak equivalence
principle
because it depends upon
the number $F$ of fermions
(minus the number of antifermions)
as $F = 3B + L$
and not upon their masses.
The potential $V_L(r)$
changes Newton's potential to
\begin{equation}
V_{NL}(r) = {} - G \, \frac{m m'}{r} 
\lt( 1 + \a \, e^{-r/\l} \rt)
\label{V(r)}
\end{equation}
in which the coupling strength $\a$ is 
\begin{equation}
\a = {} - \frac{3 F F'  f^2}{2 \pi G mm'}
= {} - \frac{3 F F' m_{\text{P}}^2 f^2}{2 \pi mm'},
\label{alpha is}
\end{equation}
and the length $\l$
is $ \l = \hbar/c m_L $. 
Couplings $\a \sim 1$ are
of gravitational strength.  
Experiments~\citep{Harris:2000zz,
Chen:2014oda,
Lee:2020zjt, Tan:2020vpf,Berge:2017ovy,Tan:2016vwu,
SQYang2012,Lee:2020zjt,Adelberger:2009zz,
Geraci:2008hb,Kapner:2006si,
Smullin:2005iv,Hoyle:2004cw,
LongChan2003,Chiaverini:2002cb,
Lee:2020zjt,Hoskins:1985tn,
Williams:2004qba,
Adelberger:2003zx,
Moody:1993ir,
Hoskins:1985tn,
Spero:1980zz,
Schlamminger:2007ht,
Decca:2005qz,Chen:2014oda,
Chiaverini:2002cb,Geraci:2008hb,
LongChan2003,Tu:2007zz,
Yang:2012zzb,Tan:2016vwu,
Fischbach:1999bc}
that test the inverse-square law and 
the weak equivalence principle
have put upper limits on the
strength
$|\a|$ of the coupling
for a wide range of lengths 
$10^{-8} < \l <  10^{13}$ m
and masses 
$2 \by 10^{-20} < m_L < 20$ eV. 
\par
Experiments that tested 
the inverse-square law
at very short distances, between
10 nm and 3 mm, were done with masses 
of gold~\citep{Harris:2000zz},
of gold and
silicon~\citep{Chen:2014oda}, of 
platinum~\citep{Lee:2020zjt}, and
of tungsten~\citep{Tan:2020vpf}.
For a mass $m$ 
of $N$ atoms of gold which has 
$F_{\text{Au}} = 670$ fermions
(quarks and electrons)
in each atom of mass 
$m_{\text{Au}} = 196.966$ u,
the ratio 
$F m_{\text{P}}/m$ that appears
in the coupling $\a$ (\ref{alpha is})
is 
\begin{equation}
\frac{N F_{\text{Au}} \, m_{\text{P}}}
{N m_{\text{Au}}}
={}
\frac{F_{\text{Au}} \, m_{\text{P}}}
{m_{\text{Au}}} = {} 
\frac{670 \, m_{\text{P}}}{196.966 \, u}
= 4.458 \by 10^{19}.
\label{gold's ratio}
\end{equation}
So the coupling 
strength is 
$\a_{\text{Au}} ={} - 9.490
\by 10^{38} \, f^2$ for gold.
An atom of silicon has
$F_{\text{Si}} = 98$ fermions
and a mass of
$m_{\text{Si}} = 28.085$\,u,
so 
$F_{\text{Si}} \, m_{\text{P}}/m_{\text{Si}} 
= 4.573$ and 
$\a_{\text{Si}} ={} - 9.988 
\by 10^{38} \, f^2$.
Platinum 
has $F_{\text{Pt}} = 663$ and
$m_{\text{Pt}} = 195.084 \, u$, 
so $\a_{\text{Pt}} = {} 
- 9.474 \by 10^{38} \, f^2$. 
Tungsten has
$F_{\text{W}} = 626$ and
$m_{\text{W}} = 183.84 \, u$, 
so
$\a_{\text{W}} = {} 
- 9.510 \by 10^{38} \, f^2$.
For such test masses,
$f^2 \approx |\a| \by 10^{-39}$. 
\par
\begin{figure}[htbp]
\begin{center}
\includegraphics[trim={5mm 7mm 0 7mm},clip,
width=0.55\textwidth]{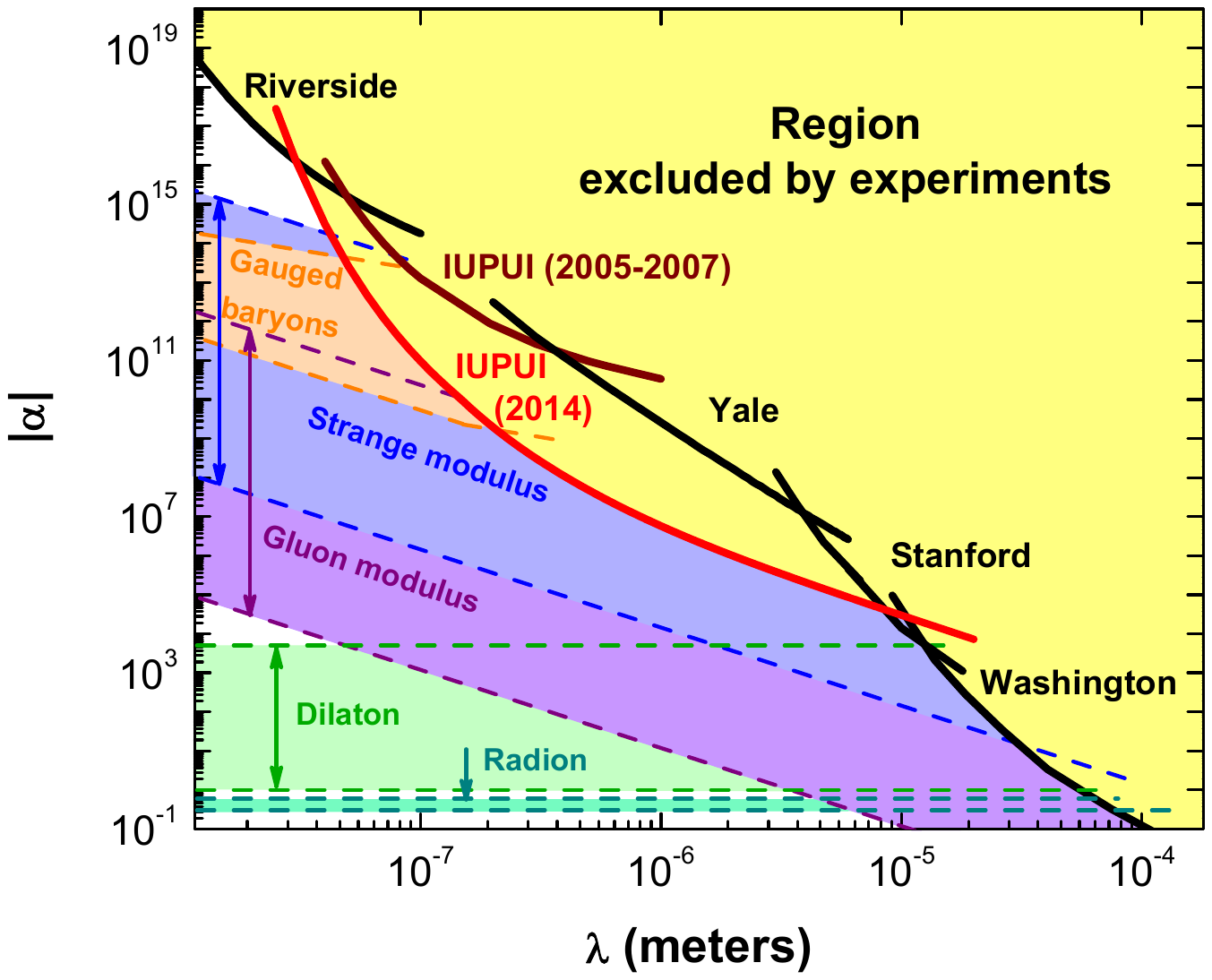}
\caption{Upper limits (95\% confidence)
on the strength $|\a_{\text{Au}}|$
of Yukawa potentials that violate
the inverse-square law at
distances $10^{-8} < \l < 2 \by 10^{-4}$ m.
(Fig. 4 of Chen et al.~\citep{Chen:2014oda})}
\label{fig 1 Chen Krause}
\end{center}
\end{figure}

The Riverside group~\citep{Harris:2000zz}
placed on the
strength $|\a_{\text{Au}}|$
an upper limit
(95\% confidence) 
that drops from 
$|\a_{\text{Au}}| \lesssim 10^{19}$
to 
$|\a_{\text{Au}}| \lesssim 10^{16}$
as the length $\l$ rises
from $10^{-8}$ m to $4 \by 10^{-8}$ m.
The IUPUI group~\citep{Chen:2014oda}
put an upper limit
(95\% confidence) on the
strength $|\a_{\text{Au-Si}}|$
that drops from 
$|\a_{\text{Au-Si}}| \lesssim 10^{16}$
to 
$|\a_{\text{Au-Si}}| \lesssim 10^{5}$
as the length $\l$ rises
from $4 \by 10^{-8}$ m to $8 \by 10^{-6}$ m.
These results of the Riverside 
and IUIPUI groups are plotted
in Fig.~\ref{fig 1 Chen Krause}
from Chen et al.~\citep{Chen:2014oda}.
\par
\begin{figure}[htbp]
\begin{center}
\includegraphics[trim={13mm 4mm 0 0 },clip,
width=0.48\textwidth]{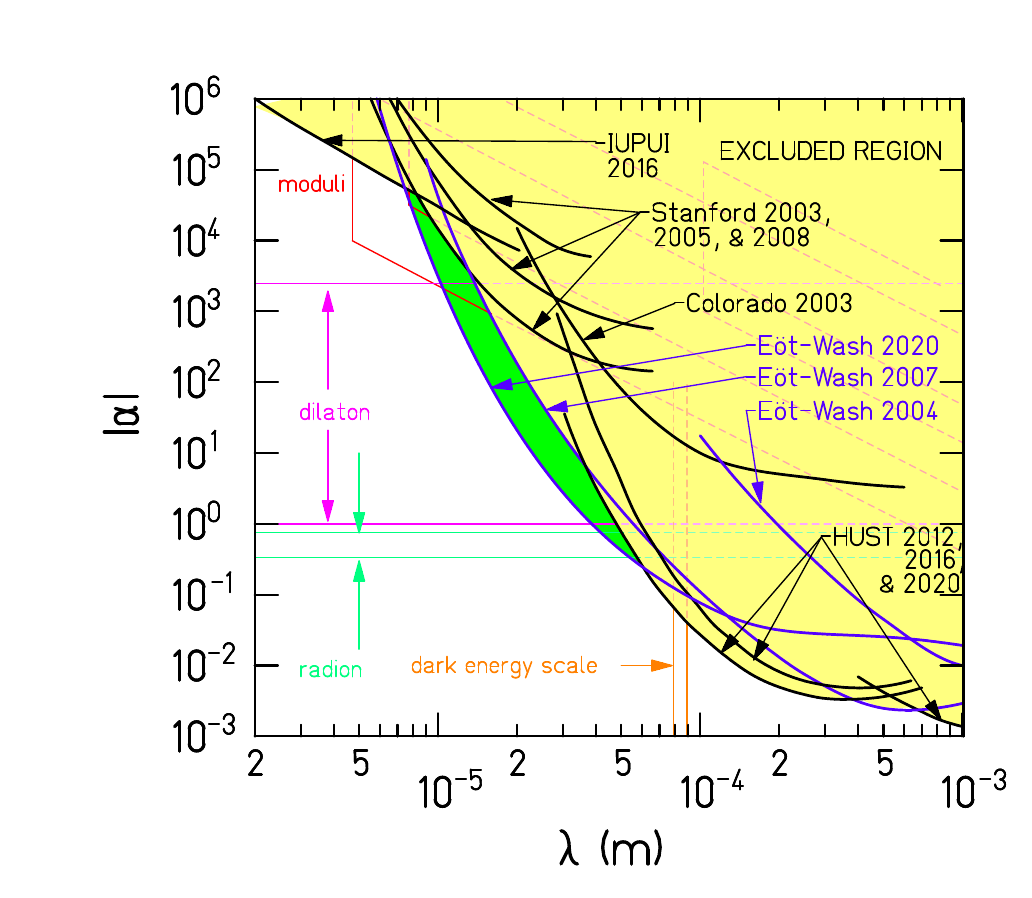}
\caption{Upper limits (95\% confidence)
on the strength $|\a_{\text{Pt}}|$ of
Yukawa potentials that violate
the inverse-square law at
sub-mm distances~\citep{Lee:2020zjt,
Tan:2020vpf,Berge:2017ovy,Tan:2016vwu,
SQYang2012,Lee:2020zjt,
Adelberger:2009zz,Geraci:2008hb,
Kapner:2006si,Smullin:2005iv,Hoyle:2004cw,
LongChan2003,Chiaverini:2002cb,
Hoskins:1985tn}.
(Fig.~5b of Lee et al.~\citep{Lee:2020zjt})
}
\label{fig 2 Lee Adelberger}
\end{center}
\end{figure}
\begin{figure}[htbp]
\begin{center}
\includegraphics[trim={5mm 0 10mm 0 },clip,
angle=270,
width=0.48\textwidth]{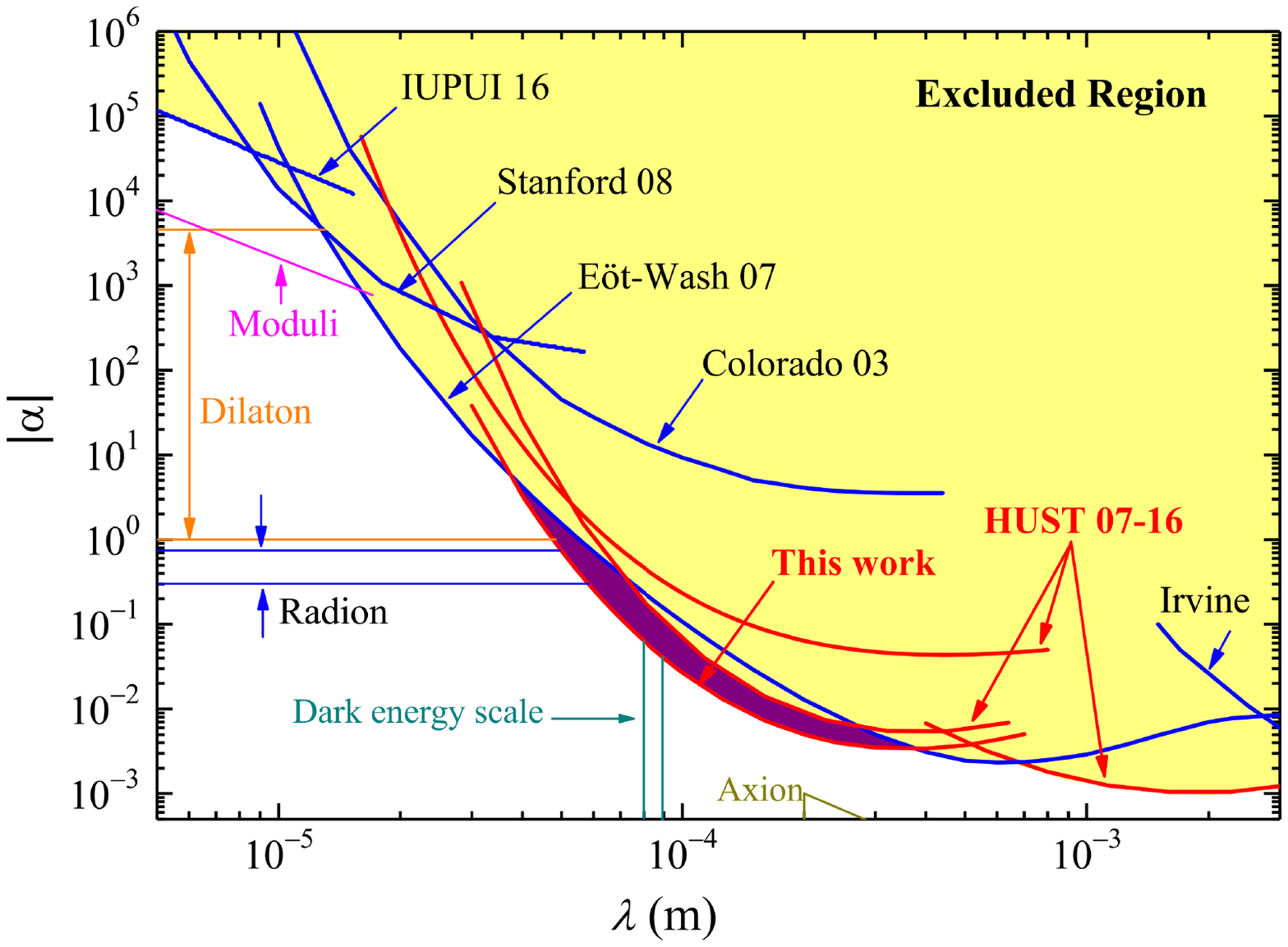}
\caption{Upper limits (95\% confidence)
on the strength $|\a_{\text{W}}|$ of
Yukawa potentials that violate
the inverse-square law at
mm and sub-mm 
distances~\citep{Tan:2020vpf,
Hoyle:2004cw,Kapner:2006si,Hoskins:1985tn,
Decca:2005qz,Chen:2014oda,
Chiaverini:2002cb,Geraci:2008hb,
LongChan2003,Tu:2007zz,
Yang:2012zzb,Tan:2016vwu}.
Light lines are theory~\citep{Adelberger:2009zz,
Adelberger:2003zx}
(Fig.~6 of Tan et al.~\citep{Tan:2020vpf}).}
\label{fig 3 Tan Du}
\end{center}
\end{figure}

Other short-distance
experiments~\citep{Lee:2020zjt, 
Tan:2020vpf,
Tan:2016vwu,
SQYang2012,
Adelberger:2009zz,Geraci:2008hb,
Kapner:2006si,Smullin:2005iv,Hoyle:2004cw,
LongChan2003,Chiaverini:2002cb,
Hoskins:1985tn,
Decca:2005qz,
Chiaverini:2002cb,Geraci:2008hb,
LongChan2003,Tu:2007zz,
Yang:2012zzb,Tan:2016vwu} 
have tested the inverse-square law 
at the slightly longer
distances of
$2 \by 10^{-6} < \l < 3 \by 10^{-3}$ m.
The Washington group~\citep{Lee:2020zjt} 
used test masses of platinum. 
The ~\citep{Tan:2020vpf}
used test masses of tungsten. 
The upper limits 
(95\% confidence) on
the strength $|\a|$ are shown
for platinum
in Fig.~\ref{fig 2 Lee Adelberger}
from Lee et al.~\citep{Lee:2020zjt}
and for tungsten
in Fig.~\ref{fig 3 Tan Du}
from Tan et al.~\citep{Tan:2020vpf}.
The upper limit on 
the strength $|\a|$ falls from 
$|\a| \lesssim 10^6$ at $ \l \sim 2 \by 10^{-6}$ m
to $|\a| \lesssim 10^4$ at $ \l \sim 8 \by 10^{-6}$ m
and then from $|\a| \lesssim 10^3$ 
at $\l \sim 10^{-5}$ m
to $|\a| \lesssim 1$ at $ \l \sim 4 \by 10^{-5}$ m
and to $|\a| \lesssim 10^{-3}$ 
at $ \l \sim 2 \by 10^{-3}$ m.

\par
Other groups~\citep{Yang:2012zzb,
Adelberger:2009zz,
Schlamminger:2007ht,
Williams:2004qba,
Adelberger:2003zx,
Moody:1993ir,
Hoskins:1985tn,
Spero:1980zz,
Fischbach:1999bc} 
have tested the
inverse-square law over a huge range
of longer distances,
$10^{-3} < \l < 3 \by 10^{15}$ m.
In 2012 the HUST group~\citep{Yang:2012zzb}
put an upper limit of 
$|\a| \lesssim 10^{-3}$ for
$7 \by 10^{-4} < \l < 5 \by 10^{-3}$ m,
while in 1985
the Irvine group~\citep{Hoskins:1985tn} 
put an upper limit
of $|\a| \lesssim 10^{-3}$ 
for lengths $7 \by 10^{-3} < \l < 10^{-1}$\,m.
\par
\begin{figure}[htbp]
\begin{center}
\includegraphics[trim={20mm 65mm 0 65mm},clip,
width=0.55\textwidth]{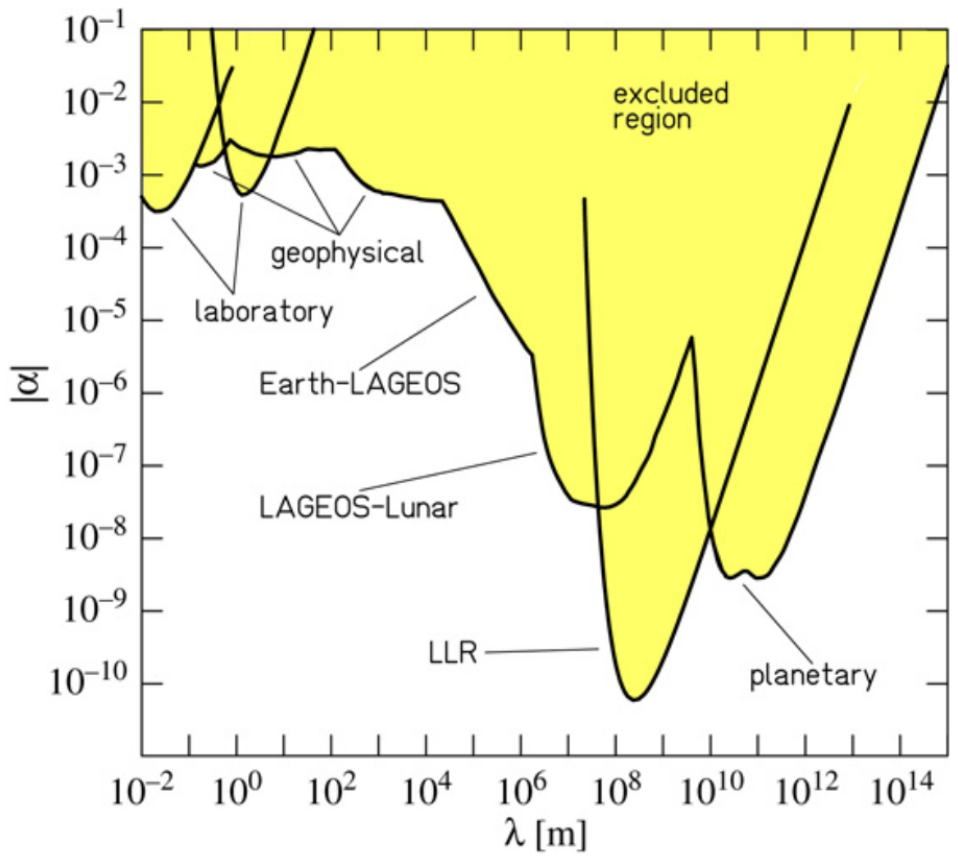}
\caption{Upper limits (95\% confidence)
on the strength $|\a|$ of Yukawa violations
of the inverse-square law at large 
distances $10^{-2} < \l < 10^{15}$ 
m~\citep{Moody:1993ir,
Hoskins:1985tn,
Spero:1980zz,Fischbach:1999bc,
Adelberger:2009zz}
(Fig.~10 of Adelberger et al.~\citep{Adelberger:2009zz}).}
\label{fig 4 Adelberger}
\end{center}
\end{figure}

Fischbach and Talmadge~\citep{Fischbach:1999bc}
and Adelberger et al.~\citep{Adelberger:2009zz}
have reported tests of the
inverse-square law for distances
in the range
$10^{-2} < \l < 10^{15}
$\,m~\citep{Adelberger:2009zz,
Moody:1993ir,
Hoskins:1985tn,
Spero:1980zz,Fischbach:1999bc}.
As shown in Fig.~\ref{fig 4 Adelberger}
from Adelberger et al.~\citep{Adelberger:2009zz},
the upper limit lies between
$|\a| < 3 \by 10^{-4}$ and
$|\a| < 2 \by 10^{-3}$ 
for 
$10^{-2} < \l < 10^4$\,m
but drops from $|\a| \lesssim 10^{-4}$ to
$|\a| \lesssim 10^{-10}$ as the length increases
from $10^4$ to $10^8$\,m.
The upper limit is about
$|\a| \lesssim 5 \by 10^{-9}$ on planetary
scales $10^{10} < \l < 5 \by 10^{11}$\,m.
\par
\begin{figure}[htbp]
\begin{center}
\includegraphics[
width=0.50\textwidth]
{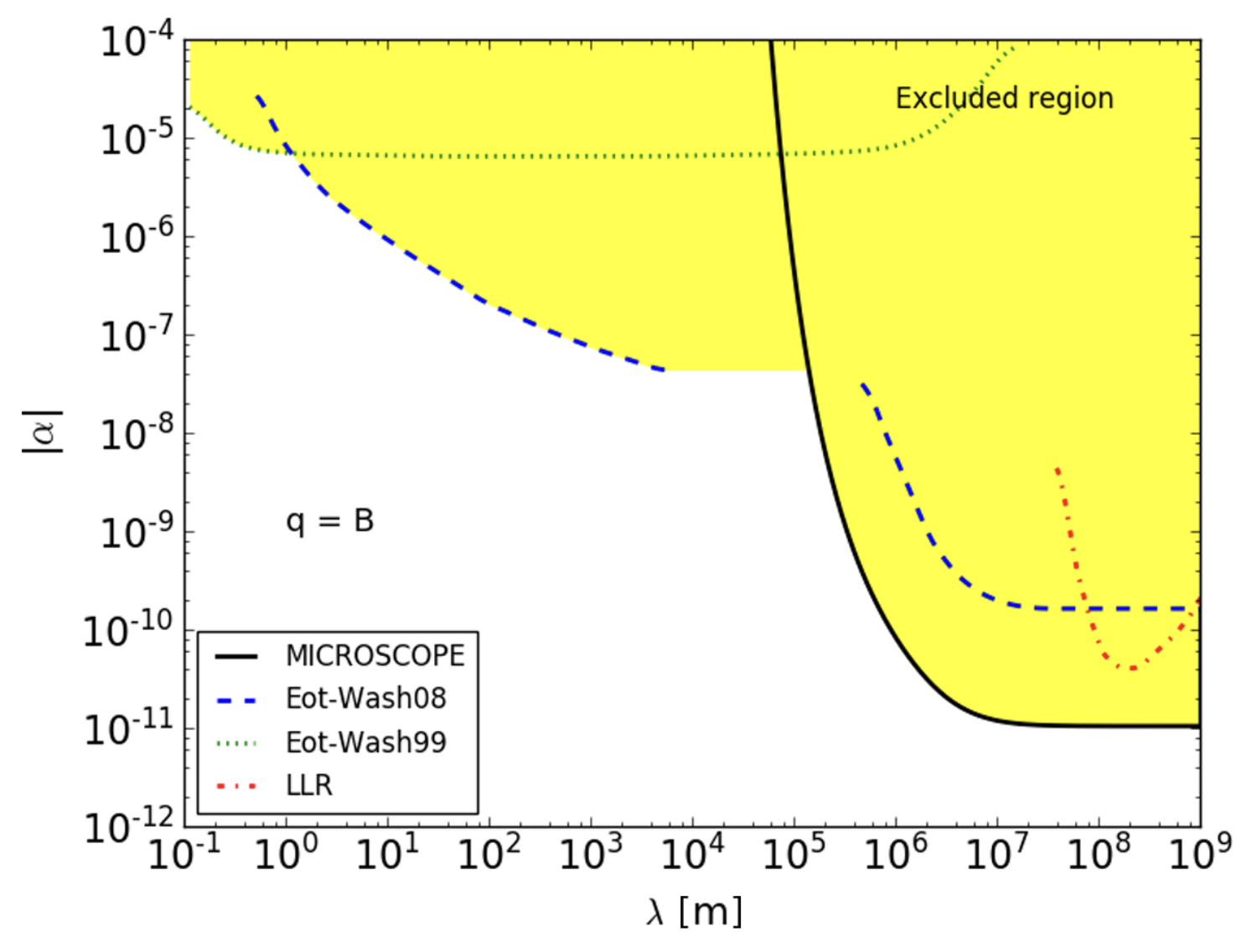}
\caption{Upper limits 
(95\% confidence)
on the strength $|\a|$ 
of Yukawa potentials that violate
the weak equivalence
principle at long
distances~\citep{Berge:2017ovy,
Adelberger:2009zz,
Williams:2004qba,
Moody:1993ir,
Hoskins:1985tn,
Spero:1980zz}
(Fig.~1 of Berg{\'{e}} et 
al.~\citep{Berge:2017ovy}).
}
\label{fig 5 Berge}
\end{center}
\end{figure}

The Washington group have used
torsion-balance experiments to
look for Yukawa potentials
that violate the weak equivalence 
principle in the range 
of distances
$0.3 < \l < 10^9
$\,m~\citep{Adelberger:2009zz}.
They have put upper limits
(95\% confidence)
on the strength $|\a|$
of the coupling to $B$,
$Z$, and $N \equiv B - L$  
but not explicitly on the coupling
to fermion number $F = 3B + L$.
For $B$, their upper limit runs from
$|\a| \lesssim 10^{-5}$ at $10^{-1}$\,m
to $|\a| \lesssim 6 \by 10^{-8}$ 
at $7 \by 10^5$\,m
and then falls to $|\a| \lesssim 10^{-10}$
for $10^7 < \l < 10^{13}$\,m
as shown by the dashed lines
in Fig.~\ref{fig 5 Berge} 
from Berg{\'{e}} et 
al.~\citep{Berge:2017ovy}.
For $Z$ and $N$, their upper limit 
runs from
$|\a| \lesssim 10^{-6}$ at $10^{-1}$\,m
to $|\a| \lesssim 6 \by 10^{-9}$ at $10^6$\,m
and then falls to $|\a| \lesssim 2 \by 10^{-11}$
for $10^7 < \l < 10^{9}$\,m~\citep{Adelberger:2009zz}.
\par
More recent satellite measurements
by the MICROSCOPE mission
have lowered the upper limit on 
the strength $|\a|$ 
of Yukawa potentials 
that violate the weak equivalence principle
by about an order of magnitude for
$10^7 < \l < 10^9$\,m~\citep{Berge:2017ovy}.
The upper limit for coupling to $B$
is $|\a| \lesssim 10^{-11}$
for $10^7 < \l < 10^9$\,m
as shown in Fig.~\ref{fig 5 Berge}
from Berg{\'e}  
et al.~\citep{Berge:2017ovy}.
Their limit for coupling to $N$ is even lower:
$|\a| \lesssim 4 \by 10^{-13}$ for 
$10^7 < \l < 10^9
$\,m~\citep{Berge:2017ovy}.

\begin{figure}[htbp]
\begin{center}
\includegraphics[trim={13mm, 65mm 0 74mm},clip,
width=0.54\textwidth] 
{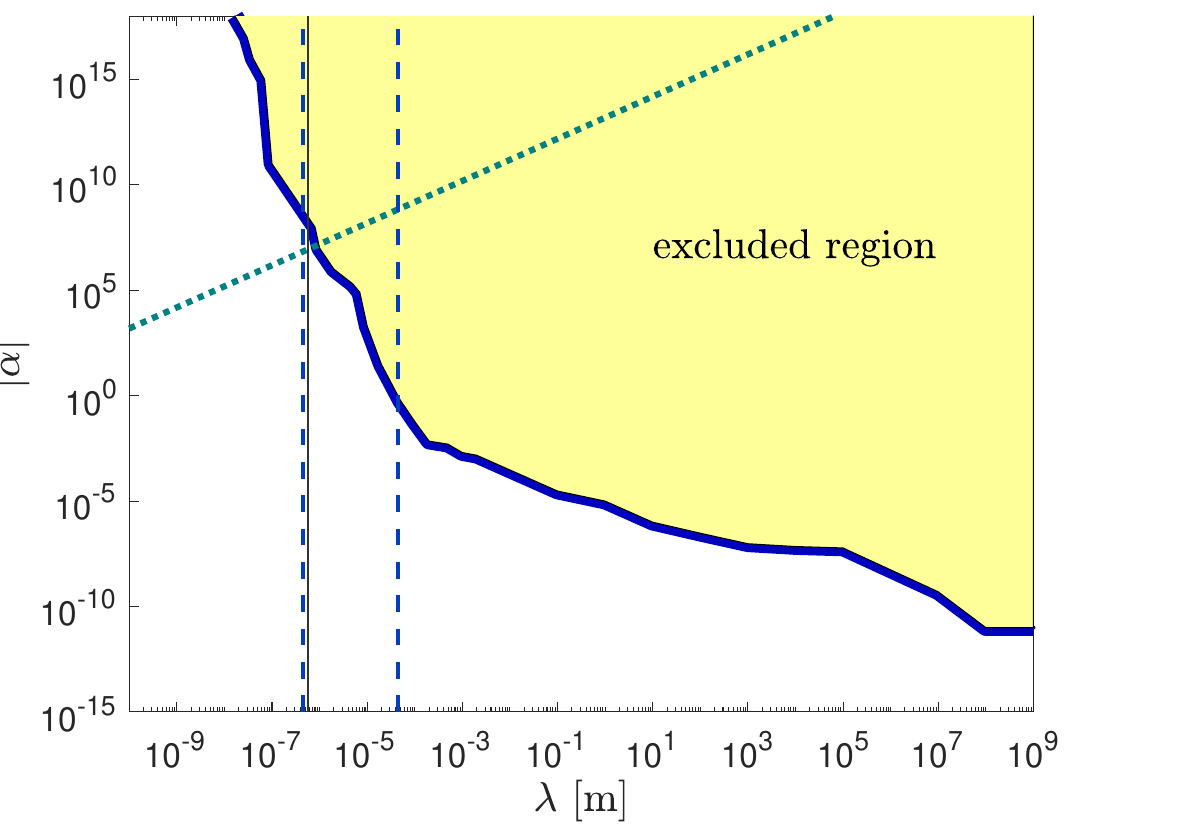}
\caption{The upper bound
(95\% confidence)
on the strength $|\a|$ 
of Yukawa potentials that violate
the inverse-square law or
the weak equivalence
principle at various
distances $\l$ is the
solid dark-blue curve~\citep{Harris:2000zz,
Chen:2014oda,
Lee:2020zjt, Tan:2020vpf,Berge:2017ovy,Tan:2016vwu,
SQYang2012,Lee:2020zjt,Adelberger:2009zz,
Geraci:2008hb,Kapner:2006si,
Smullin:2005iv,Hoyle:2004cw,
LongChan2003,Chiaverini:2002cb,
Lee:2020zjt,Hoskins:1985tn,
Williams:2004qba,
Adelberger:2003zx,
Moody:1993ir,
Hoskins:1985tn,
Spero:1980zz,
Schlamminger:2007ht,
Decca:2005qz,Chen:2014oda,
Chiaverini:2002cb,Geraci:2008hb,
LongChan2003,Tu:2007zz,
Yang:2012zzb,Tan:2016vwu,
Fischbach:1999bc}.
The region under the dotted
green line denotes $L$ bosons
with lifetimes greater than the
age of the universe for the 
case in which the fudge factor 
$b = 1$.
The region below that
line and between the vertical
dashed blue lines denotes $L$ bosons
that are between 1 and 100\%
of dark matter.
The thin vertical gray solid line 
marks the wavelength of an $L$ boson
whose mass is 
$2 \, m_{\nu_e}^{\text{eff}}
={}  2.2$ eV~\citep{ZylaPDG2020}.
}
\label {fig 6 me}
\end{center}
\end{figure}

\par
Some  of these important 
results~\citep{Harris:2000zz,
Chen:2014oda,
Lee:2020zjt, Tan:2020vpf,Berge:2017ovy,Tan:2016vwu,
SQYang2012,Lee:2020zjt,Adelberger:2009zz,
Geraci:2008hb,Kapner:2006si,
Smullin:2005iv,Hoyle:2004cw,
LongChan2003,Chiaverini:2002cb,
Lee:2020zjt,Hoskins:1985tn,
Williams:2004qba,
Adelberger:2003zx,
Moody:1993ir,
Hoskins:1985tn,
Spero:1980zz,
Schlamminger:2007ht,
Decca:2005qz,Chen:2014oda,
Chiaverini:2002cb,Geraci:2008hb,
LongChan2003,Tu:2007zz,
Yang:2012zzb,Tan:2016vwu,
Fischbach:1999bc}
are summarized 
in broad-brush fashion
in Fig.~\ref{fig 6 me}.
The upper bound 
(95\% confidence) on $|\a|$
is the solid dark-blue curve
which
falls from $10^{19}$ for 
$\l = 7 \by 10^{-8}$ m
to $10^{-11}$ at $\l = 10^{9}$ m.
The $(\l, |\a|)$ region above this curve
is excluded.
Points in the allowed region
that are below the blue-green dotted
line correspond to $L$ bosons
with lifetimes longer than the age 
of the universe.
Those that also are 
between the vertical dashed
lines denote effectively stable $L$ bosons 
whose masses could account for
between 1 and 100\%
of dark matter.

\section{Lorentz bosons as dark matter}
\label{Lorentz bosons as dark matter sec}

\par
Analysis of the double galaxy cluster
1E0657-558 (the ``bullet cluster" 
at $z=0.296$) 
suggests~\citep{Clowe:2006eq, Weinberg2010p186} 
that dark matter 
interacts weakly, perhaps
with gravitational strength
$|\a| \sim 1$.
As of now, there has been no 
accepted detection of dark matter 
in a laboratory.
\par
The experiments~\citep{Harris:2000zz,
Chen:2014oda,
Lee:2020zjt, Tan:2020vpf,Berge:2017ovy,Tan:2016vwu,
SQYang2012,Lee:2020zjt,Adelberger:2009zz,
Geraci:2008hb,Kapner:2006si,
Smullin:2005iv,Hoyle:2004cw,
LongChan2003,Chiaverini:2002cb,
Lee:2020zjt,Hoskins:1985tn,
Williams:2004qba,
Adelberger:2003zx,
Moody:1993ir,
Hoskins:1985tn,
Spero:1980zz,
Schlamminger:2007ht,
Decca:2005qz,Chen:2014oda,
Chiaverini:2002cb,Geraci:2008hb,
LongChan2003,Tu:2007zz,
Yang:2012zzb,Tan:2016vwu,
Fischbach:1999bc} 
sketched in 
Sec.~\ref{Tests of the Inverse-Square Law sec}
put no upper limits on the mass
$m_L = h/c \l$ of $L$ bosons
and no lower
limits on their coupling $|\a|$.
The proposed $L$ bosons 
are electrically neutral.
If their mass is heavy enough and
if their coupling is sufficiently weak,
then they would be an effectively stable
part of dark matter.
\par
Because they couple to fermion number
and not to mass, their coupling $f^2$ is 
much weaker than $|\a|$ by a factor
related to Avogadro's number.
For the metals 
(Au, Si, Pt, and W)
used in many of the
experiments~\citep{Harris:2000zz,
Chen:2014oda,
Lee:2020zjt, Tan:2020vpf,Berge:2017ovy,Tan:2016vwu,
SQYang2012,Lee:2020zjt,Adelberger:2009zz,
Geraci:2008hb,Kapner:2006si,
Smullin:2005iv,Hoyle:2004cw,
LongChan2003,Chiaverini:2002cb,
Lee:2020zjt,Hoskins:1985tn,
Williams:2004qba,
Adelberger:2003zx,
Moody:1993ir,
Hoskins:1985tn,
Spero:1980zz,
Schlamminger:2007ht,
Decca:2005qz,Chen:2014oda,
Chiaverini:2002cb,Geraci:2008hb,
LongChan2003,Tu:2007zz,
Yang:2012zzb,Tan:2016vwu},
the relation is 
\begin{equation}
f^2 \sim |\a| \by 10^{-39}.
\label{for Au the relation is}
\end{equation}
Even for the highest upper limit
$|\a| < 10^{19}$
shown in Fig.~\ref{fig 6 me},
the coupling of the
$L$ bosons is only
$f^2 \lesssim 10^{-20}$.
\par
Because they interact so weakly,
$L$ bosons decay slowly.
The decay width of the $Z$ boson is 
$\Gamma_Z = 3.7 \, e^2 \, m_Z c^2/4\pi 
= 2.5$  
GeV, and its lifetime is
$\tau_Z = \hbar/\Gamma_Z 
= 2.6 \by 10^{-25}$\,s. 
The analog of the
electromagnetic coupling
$e^2/4\pi \sim 1/137$
for $L$ bosons
is $f^2/4\pi$.
In terms of $f^2$ and $|\a|$
(\ref{for Au the relation is}),
the decay width of an 
$L$ boson of mass
$m_L$ is roughly
\begin{equation}
\Gamma_L \sim {} \frac{b \, f^2 \, 
m_L c^2}{4\pi} = 
\frac{b \, |\a| \, 
m_L c^2}{4\pi}
\by 10^{-39} 
\end{equation}
in which $b$ 
is a fudge factor,
$0.1 \lesssim b \lesssim 10$, that
depends on the decay channels.
The $L$ boson lifetime then is
\begin{equation}
\begin{split}
\tau ={}& 
\frac{\hbar}{\Gamma_L}
= \frac{8.3 \by 10^{24}} 
{b \, |\a| \, m_L c^2[\text{eV}]} \, 
\, \text{s}
=
\frac{1.9}{b|\a|} \,
\frac{10^7}{m_Lc^2[\text{eV}]}
\,\,   t_0
\label{lifetime of L boson}
\end{split}
\end{equation}
in which 
$t_0 = 4.356 \by 10^{17}$s 
is 13.8 billion years,
the age of the universe.
An $L$ boson of mass
$m_L < (19/b|\a|) \, \text{MeV}$ 
is effectively stable in 
that its lifetime exceeds the
age of the universe.
If the fudge factor $b$
is taken to be unity,
then $L$ bosons of wavelength $\l$
are effectively stable for couplings
\begin{equation}
|\a| \lesssim 
\lt( 1.5 \by 10^{13} \rt) 
\,\, \l \, [\text{m}]
\label{alpha vs lambda}
\end{equation}
which is the dotted green line
in Fig.~\ref{fig 6 me}.
Points below it denote effectively
stable $L$ bosons.
\par
If the lightest fermion has mass
$m_{\text{lightest}}$, then
$L$ bosons of mass
less than $2\,m_{\text{lightest}}$ would
be absolutely stable.
The dash-dotted gray vertical
line in Fig.~\ref{fig 6 me} 
is the wavelength 
$ \l = 5.6 \by 10^{-7}$m
of twice the upper limit 
on the effective mass of 
the electron neutrino,
$2 \, m_{\nu_e}^{(\text{eff})}
$~\citep{ZylaPDG2020}.
\par
The mass density of cold dark matter is
$\rho_{\text{cdm}} =
(2.2414 \pm 0.017) \by 10^{-27}$
kg m$^{-3}$~\citep[col. 7, p. 15]
{Aghanim:2018eyx}.
So if all of cold dark matter
were made of $L$ bosons 
of mass $m_L$, then
their number density would be
\begin{equation}
\begin{split}
n_L 
={}& 
\frac{\rho_{\text{cdm}}} 
{m_L}
={} 
\frac{1.26 \by 10^{9} } 
{m_L c^2 [\text{eV}]} \, \text{m}^{-3}.
\end{split}
\end{equation}
\par
To estimate the present
number density of each kind
of $L$ boson, I'll assume
that the $L$ bosons are
effectively stable and 
have not interacted since 
they dropped out of equilibrium
in the very early universe.
\par
At temperaures
$ kT \gg m_L c^2$
so high that the weakly interacting
$L$ bosons were in 
thermal equilibrium,
the number density of
each 
of the six $L$ bosons is
given by the Planck distribution as 
\begin{equation}
n(T) ={}
\frac{3 \zeta(3) (kT)^3}
{\pi^2 (\hbar c)^3}
= \frac{9.609 \by 10^7 \, T^3} 
{(\text{m\,K})^3} .
\label{number density}
\end{equation}
\par
In the limit of vanishing
coupling $|\a| \to 0$,
the number $n(t) a^3(t)$
of $L$ bosons 
within a fixed comoving box 
does not change with time.
So the number now
$n(t_0) a^3(t_0) = n(t_0)$
is the number at any earlier 
time multiplied by $a^3(t)$
\begin{equation}
n(t_0) = n(t) a^3(t).
\end{equation}
At \tit{very} early times,
we may approximate
the integral for the time
as a function of the scale factor 
$a$ as~\citep{Cahillp518}
\begin{align}
t(a) = {}& \frac{1}{H_0} \int_0^{a}
\!\! \frac{dx}{ \sqrt{\Omega_{\Lambda} \, x^2
+ \Omega_{k}  
+ \Omega_{m} \, x^{-1}  
+ \Omega_{r} \, x^{-2} } } 
\nn \\
\approx {}&
\frac{1}{H_0} \int_0^{a}
\!\! \frac{x \, dx}{ \sqrt{ \Omega_{r} } } 
= \frac{a^2}{2 H_0 \sqrt{ \Omega_{r} }}.
\label {t(x)}
\end{align}
The Hubble constant and 
the fraction
$\Omega_r = 9.0824 \by 10^{-5}$
then give us the scale-factor as
\begin{equation}
a(t) = 
\sqrt{2H_0 \sqrt{\Omega_r} \, t}
=
2.04 \by 10^{-10} \sqrt{t \text[s]}.
 \label{scale-factor}
\end{equation} 
\par 
If $\mcl{N}$ types of particles 
made up the radiation 
at very early times, then the time
and the temperature were 
related by~\citep{Weinberg2010p152}
\begin{equation}
\sqrt{\mcl{N}} \,\, t \, T^2
={}
\sqrt{\frac{3c^2}
{16 \pi G a_{\text{r}}}} 
= 3.25924 \by 10^{20} \,\, \text{s\,K}^2
\label{time-temperature relation}
\end{equation}
in which $a_{\text{r}}$ is the
radiation constant
\begin{equation}
a_{\text{r}} = {} 
\frac{\pi^2 k^4}{15 \hbar^3 c^3}
=
7.56577(5) \by 10^{-16} \,
\text{J m}^{-3} \, \text{K}^{-4}.
\label {radiation constant}
\end{equation}
\par
So in terms of the number density
(\ref{number density}),
the scale-factor
(\ref{scale-factor}),
and the time-temperature relation
(\ref{time-temperature relation}),
the number density is roughly
\begin{equation}
\begin{split}
n(t_0) = {}
n(t) a^3(t)
=
\frac{4.8 \by 10^9 }
{\mcl{N}^{3/4}} \, \text{m}^{-3}.
\label{L density}
\end{split}
\end{equation}
In the standard model, 
$\mcl{N} = 126$, but 
the actual number 
relevant at high temperatures
may be much higher.
If we assume that $\mcl{N} = 4^4$,
then $\mcl{N}^{-3/4} = 1/64$,
and the present number density of each
kind of $L$ boson would be
\begin{equation}
n(t_0) =
7.5 \by 10^7 \, \text{m}^{-3}.
\label{L density 64}
\end{equation}
\par
Let us further assume that
all six $L$ bosons get the same 
mass $m_{L}$.
In this case, if their
mass density is not to 
exceed the density of dark matter,
then the inequality 
\begin{equation}
6 \, m_{L} \, n(t_0) < \rho_{\text{cdm}}
= 2.24 \by 10^{-27} \, \text{kg m}^{-3}
\end{equation}
implies that the mass $m_L$
must be less than
\begin{equation}
\begin{split}
m_{L} < 4.9 \by 10^{-36} \, \text{kg} 
={}& 2.8 \, \text{eV}/c^2.
\label{limits on L boson masses}
\end{split}
\end{equation}
The lifetime of an $L$ boson 
(\ref{lifetime of L boson})
would then be
\begin{equation}
\tau_{L} > \frac{3.4}{b|\a|} 
\by 10^6 \, t_0 
\end{equation}
which for $b |\a| < 1.7 \by 10^6$
exceeds the age
$t_0$ of the universe. 
The range 
$\l_{L} = h /m_{L} c$ of the
corresponding Yukawa potential 
is
\begin{equation}
\l_{L} > 4.5 \by 10^{-7} \, \text{m}.
\end{equation}
Points $(\l,|\a|)$ below the dotted green line
in Fig.~\ref{fig 6 me} and between its 
vertical dashed blue-green lines denote
$L$ bosons constituting between
1 and 100\% of the dark matter.
The upper limit on the effective mass
of the electron neutrino is
$m_{\nu_e}^{(\text{eff})} < 1.1$\,eV~\citep{ZylaPDG2020}.
The thin gray vertical line labels
$L$ bosons of mass 
$m_L = 2 \, m_{\nu_e}^{\text{eff}}$.

\section{Conclusions}
\label{Conclusions sec}

\par
General relativity
with fermions has two
independent symmetries:
general coordinate
invariance and
local Lorentz invariance.
General general coordinate
invariance acts on coordinates
and on the world indexes
of tensors but leaves Dirac
and Lorentz indexes
unchanged.
Local Lorentz invariance
acts on Dirac
and Lorentz indexes but
leaves world indexes and
coordinates unchanged.
It acts like an internal symmetry.
\par
General coordinate
invariance is implemented
by the Levi-Civita connection
$\Gamma^j_{\ph{j} k i}$ 
and by 
Cartan's tetrads $c^a_{\ph{a} i}$.
In the standard formulation
of general relativity
with fermions,
local Lorentz invariance
is implemented by the
same fields in a combination 
called the
spin connection
$\om^{a b}_{\ph{a b} \, i }
= c^a_{\ph{a} j} \, c^{b k} 
\, \Gamma^j_{\ph{j} k i}
+  c^a_{\ph{a} k} \, \p_i c^{b k}$.
These fields all have the same action,
the Einstein-Hilbert action $R$.
\par
Because local Lorentz invariance
is different from and independent 
of general coordinate
invariance,
it is suggested in this paper
that local Lorentz invariance
is implemented 
by different and independent fields
$L^{a b}_{\ph{a b} \, i }$
that gauge the Lorentz group
and that have their own 
Yang-Mills-like action.
\par
The replacement of the spin connection 
with Lorentz bosons moves 
general relativity closer to 
gauge theory and simplifies the
standard covariant derivative 
\begin{equation}
\Big( \p_i + \teighth         
\big( c^a_{\ph{a} j} \, c^{b k}
\Gamma^j_{\ph{j} k i} 
+  c^a_{\ph{a} k} \, \p_i c^{b k}
 \big)
\bos [ 
\gamma_a, \gamma_b \bos ]
\Big) \psi 
\label{standard covariant derivative}
\end{equation}
to
\begin{equation}
\lt(\p_i + \teighth 
\, 
L^{ a b }_{\ph{a b} i}
\, \bos{[} \c_a, \c_b \bos ] 
\rt) \psi .
\label{new covariant derivative}
\end{equation}
Whether the Dirac action has 
the spin-connection form
(\ref{standard covariant derivative})
or the Lorentz-boson form
(\ref{new covariant derivative})
is an experimental question.
\par
Because the proposed action
(\ref{new Dirac action II})
couples the gauge fields 
$L^{ a b }_{\ph{a b} i}$ to
fermion number and not to mass,
it violates the
weak equivalence principle.
It also leads to a Yukawa potential
(\ref{would have a static potential})
that violates
Newton's inverse-square law.
\par
Experiments~\citep{Harris:2000zz,
Chen:2014oda,
Lee:2020zjt, Tan:2020vpf,Berge:2017ovy,Tan:2016vwu,
SQYang2012,Lee:2020zjt,Adelberger:2009zz,
Geraci:2008hb,Kapner:2006si,
Smullin:2005iv,Hoyle:2004cw,
LongChan2003,Chiaverini:2002cb,
Lee:2020zjt,Hoskins:1985tn,
Williams:2004qba,
Adelberger:2003zx,
Moody:1993ir,
Hoskins:1985tn,
Spero:1980zz,
Schlamminger:2007ht,
Decca:2005qz,Chen:2014oda,
Chiaverini:2002cb,Geraci:2008hb,
LongChan2003,Tu:2007zz,
Yang:2012zzb,Tan:2016vwu,
Fischbach:1999bc}
have put upper limits on the 
strength $|\a|$ 
of the Yukawa potentials
(\ref{V(r)})
that violate
the inverse-square law and 
the weak equivalence principle
for distances 
$10^{-8} < \l < 10^9$\,m.
The upper limit ranges from
$|\a| < 10^{19}$ at $\l = 10^{-8}$\,m
to $|\a| < 10^3$ at $\l = 10^{-5}$\,m
and to 
$|\a| < 10^{-11}$ at $\l = 10^9$\,m.
There are no experimental 
lower limits on the coupling
at any distance,
so $L$ bosons 
could have lifetimes
that exceed the age of 
the universe.
There are no experimental
upper limits on the masses
of $L$ bosons.
Long lived, massive, weakly interacting,
neutral $L$ bosons would contribute
to dark matter.
From the obvious requirement
that they could make up all of
dark matter but not more,
we can infer
a crude theoretical upper limit 
on their mass of 
$m_{L} \lesssim 2.8$ eV$/c^2$
if all 6 are stable and 
have the same mass.
\par
The discovery of
a violation of the inverse-square law
by future experiments would not
be enough to 
establish the existence of 
$L$ bosons 
because the violation could be due
to the physics of a quite different
theory.
\par
If $L$ bosons are discovered, 
physicists will decide how
to think about
the force they mediate.
The force
might be considered to be 
gravitational because it arises
in a theory that is a modest
and natural extension of 
general relativity.  
But the force is not carried 
by gravitons. 
It is carried by  
$L$ bosons,
and they
implement a 
symmetry,
local Lorentz invariance,
that is
independent of 
general coordinate
invariance.
So the force
is new and might be
called a Lorentz force.

\par
\begin{acknowledgements}
I am grateful to E. Adelberger,
R. Allahverdi, 
D. Krause, E. Fischbach,
and A. Zee for helpful email.
\end{acknowledgements}

\bibliography{physics,math,lattice,books}
 
\end{document}